\documentclass[prx,amsmath,amssymb,notitlepage,twocolumn,
nofootinbib,
superscriptaddress,
longbibliography
]{revtex4-2}
\usepackage{MnSymbol}
\usepackage{amsmath}
\usepackage{tabularx,graphicx}
\usepackage{epstopdf}
\usepackage{graphicx}
\usepackage{spectralsequences} 
\usepackage{latexsym}
\usepackage{amssymb}
\usepackage{tikz-cd}
\usepackage{amsmath}
\usepackage{color, colortbl}
\usepackage{psfrag}
\usepackage{bbm}
\usepackage{bm}
\usepackage{titlesec}
\usepackage{dsfont}
\usepackage{feynmp}
\usepackage{slashed}
\usepackage{multirow}
\usepackage{framed}
\usepackage{enumitem}
\newcommand{\beq}{\begin{eqnarray}}
\newcommand{\eeq}{\end{eqnarray}}

\newcommand{\Tr}{{\rm Tr}}
\newcommand{\tr}{{\rm tr}}

\newcommand{\bsp}{\begin{split}}
\newcommand{\esp}{\end{split}}

\newcommand{\poly}{{\rm Poly}}

\newcommand{\ie}{{i.e., }}
\newcommand{\eg}{{e.g., }}

\newcommand{\cov}{{\rm{Cov}}}
\newcommand{\mO}{{\mathcal{O}}}

\newcommand{\R}{{\mathcal{R}}}
\newcommand{\E}{{\mathcal{E}}}
\newcommand{\C}{{\mathcal{C}}}
\newcommand{\mL}{{\mathcal{L}}}
\newcommand{\M}{{\mathcal{M}}}

\usepackage{braket}
\newcommand{\bbeta}{{\boldsymbol\beta}}
\newcommand{\polylog}{{\rm Polylog}}
\newcommand{\D}{{\mathcal{D}}}
\newcommand{\G}{{\mathcal{G}}}
\newcommand{\A}{{\mathcal{A}}}

\renewcommand{\S}{{\mathcal{S}}}

\newcommand{\identity}{{\mathbb{I}}}
\newcommand{\bh}{{\bf h}}
\newcommand{\rng}{{\rm rng\,}}
\newcommand{\diam}{{\rm diam}}
\newcommand{\eq}[1]{\overset{{#1}}{=}}
\newcommand{\mysim}[1]{\overset{\small{#1}}{\,\sim\,}}
\newcommand{\opnorm}[1]{\left\Vert{#1}\right\Vert}
\renewcommand{\tilde}[1]{\widetilde{#1}}
\usepackage{amsmath}
\usepackage{amssymb}
\usepackage{color}
\definecolor{darkblue}{rgb}{0.,0.,0.4}
\definecolor{darkred}{rgb}{0.5,0.,0.}
\definecolor{BlueViolet}{RGB}{138,43,226}
\definecolor{SkyBlue}{RGB}{30,144,255}
\definecolor{DarkGreen}{RGB}{0,100,0}
\usepackage[pdftex,colorlinks=true,linkcolor=darkblue,citecolor=blue,urlcolor=darkred]{hyperref}
\usepackage[normalem]{ulem}

\def\H{\mathcal{H}}

\usepackage{tabularx}
\usepackage{tikz}
\usepackage{amsthm}
\theoremstyle{plain}
\newtheorem*{theorem*}{Theorem}
\newtheorem{theorem}{Theorem}
\newtheorem{lemma}{Lemma}

\newtheorem{definition}{Definition}

\usepackage{tcolorbox}
\tcbuselibrary{skins,breakable}

\tcolorboxenvironment{theorem}{
  enhanced, breakable,
  colback=black!5,    
  colframe=black!15,  
  boxrule=0.5pt,
  sharp corners,      
  before skip=10pt, after skip=10pt,
  left=8pt, right=8pt, top=6pt, bottom=6pt
}

\usetikzlibrary{positioning,shapes.geometric}

\begin{document}
	\title{Circuit-based characterization of finite-temperature quantum phases and self-correcting quantum memory}
	
	\author{Ruochen Ma}
    \email{ruochenma@ucsb.edu}
    \affiliation{Kavli Institute for Theoretical Physics, University of California, Santa Barbara, CA 93106, USA}
    \affiliation{Department of Physics, University of California, Santa Barbara, CA 93106, USA}
    \affiliation{Kadanoff Center for Theoretical Physics \& Enrico Fermi Institute, The University of Chicago, Chicago, Illinois, USA 60637}

    \author{Vedika Khemani}
    \email{vkhemani@stanford.edu}
    \affiliation{Department of Physics, Stanford University, Stanford, California 94305, USA}

    \author{Shengqi Sang}
    \email{sangsq@stanford.edu}
    \affiliation{Department of Physics, Stanford University, Stanford, California 94305, USA}

\begin{abstract}
Quantum phases at zero temperature can be characterized as equivalence classes under local unitary transformations: two ground states within a gapped phase can be transformed into each other via a local unitary circuit. 
We generalize this circuit-based characterization of phases to systems at finite-temperature thermal equilibrium described by Gibbs states.
We construct a channel circuit that approximately transforms one Gibbs state into another provided the two are connected by a path in parameter space along which a certain correlation-decay condition holds. For finite-dimensional systems of linear size $L$ and approximation error $\epsilon$, the locality of the circuit is $\polylog(\poly(L)/\epsilon)$. 
The correlation-decay condition, which we specify, is expected to be satisfied in the interior of many noncritical thermal phases, including those displaying discrete symmetry breaking and topological order. 
As an application, we show that any system in the same thermal phase as a zero-temperature topological code coherently preserves quantum information for a macroscopically long time, establishing self-correction as a universal property of thermal phases. As part of the proof, we provide explicit encoding and decoding channel circuits to encode information into, and decode it from, a system in thermal equilibrium.

\end{abstract}

\maketitle

\tableofcontents

\section{Introduction}

Over the last several decades, significant progress has been made in understanding zero-temperature quantum phases of gapped local Hamiltonians. This progress covers both the usual symmetry-breaking phases in Landau’s picture and many exotic phases with topological or fracton order \cite{2023McGreevy,2017wenTOreview,2019fractons}. In this context, a central role is played by a circuit-based definition of phases inspired by quantum information theory: two pure states are in the same phase if there exists a shallow-depth local unitary circuit connecting them~\cite{2010chenguwen}. This definition underlies most discussions of the classification and characterization of zero-temperature quantum phases. 

Recently, thanks to improved experimental control of devices and other platforms~\cite{2018NISQ}, the study of quantum phases has shifted to open systems described by mixed states. Following the paradigm set by ground-state quantum phase studies, an analogous circuit-based definition has been proposed: two mixed states are in the same phase if they are related by a pair of local quantum channel circuits~\cite{2019Coser,2023ASPT,hastings2011topological}, hereafter referred to as two-way connectivity. Based on this, several classification schemes have been proposed \cite{2023ASPT,2024uniformity,2025sohalabhinav,2025EllisonCheng,2024chengrover1,2024chengrover2}, and concrete examples of nontrivial phases have been constructed \cite{2022openSPT,2025ASPT,2025intrinsicTO,2025SWSSB,2025higherformanomaly,2023BFVA,2024FBAV,2023LJX,lu2023mixed,2025zhangTO,2025shi,20253DfTC, sala2025decoherence,2025mixedstateanomaly}.

Gibbs states describing systems in thermal equilibrium are among the most ubiquitous mixed states in nature. It nevertheless remains unclear whether the circuit-based characterization is applicable to Gibbs states and the thermal phases formed by them, where the latter is traditionally defined as a region of parameter space in which the free energy is analytic. To recap, for ground states, the circuit connecting states within a phase is furnished by quasi-adiabatic continuation~\cite{2005hastingswen,2010hastings}, which exists whenever the energy gap does not close. For Gibbs states, an analogous circuit construction that transforms Gibbs states within the same thermal phase is lacking, and it is unclear what the counterpart of energy gap should be.

\begin{figure}
    \centering
    \includegraphics[width=0.4\linewidth]{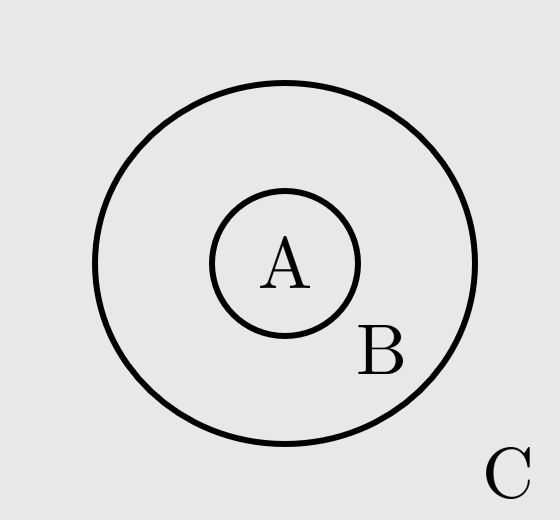}\\
    \caption{Annulus tripartition for defining clustering of correlations, illustrated in 2D. $A$ is a small region, $B$ is a region surrounding $A$, and $C$ denotes the rest of the system. 
    }\label{fig:annulus}
    \vspace{0.3cm}
    \includegraphics[width=0.75\linewidth]{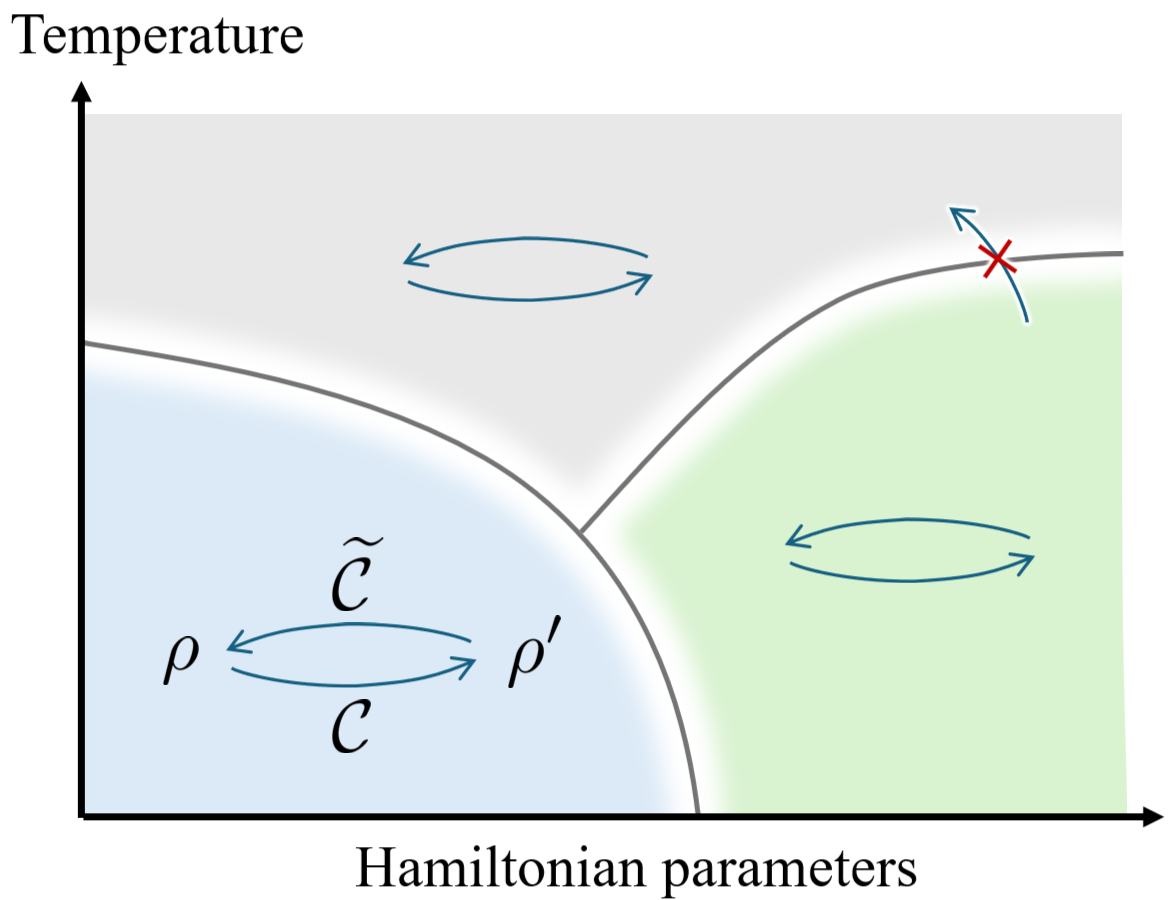}
    \caption{Sketch of three thermal phases in the temperature-parameter space, separated by phase boundaries (solid lines). In this work, we construct $\polylog(L)$-local channel circuits that transform between different Gibbs states within each phase, provided that the states in the phase satisfy a correlation-decay condition defined in Def.\ref{def:stable_clustering} (Theorem \ref{thm:connectivity}). If the Hamiltonians are commuting, then only a weaker condition, Def.\ref{def:clustering}, is needed, and the circuit is improved to a continuous-time Lindbladian (Theorem \ref{thm:cont_connectivity}). 
    {The condition is expected to be satisfied in many ordered phases, including those displaying discrete symmetry breaking and topological order, but violated when approaching phase boundaries.}
    }
     \label{fig:main}
\end{figure}

In this work, we present a general construction of local channel circuits that transform Gibbs states within the same thermal phase into one another (See Fig.\ref{fig:main} for an illustration). More specifically, for a system of linear size $L$, we construct a $\polylog(\poly L/\epsilon)$-local channel circuit that approximately transforms one Gibbs state into another with error $\epsilon$, provided that they are connected by a path in the parameter space along which a correlation-decay condition, dubbed \textbf{stable clustering} (briefly described below and formally defined in Def.\ref{def:stable_clustering}), is satisfied. 
With an additional standard assumption on the density of states at low energy~\cite{2007Has}, our construction continues to work in the zero-temperature limit for gapped Hamiltonians.

Colloquially speaking, a Gibbs state $\rho\propto e^{-H}$ satisfies stable clustering if its neighboring Gibbs states of the form $\rho'\propto e^{-H-\delta H}$ all display clustering of correlations when $\delta H$ is a small perturbation that is a sum of local terms (temperature is absorbed into $H$ and $\delta H$). Here, by `clustering of correlations' we mean that the connected correlation function between any local observable $O_1$, supported in region $A$ in Fig.\ref{fig:annulus}, and any other, possibly nonlocal observable $O_2$, supported in region $C$, decays exponentially with the distance between their spatial supports. If the Hamiltonians under consideration satisfy a global (0-form) symmetry, then $O_1$ and $O_2$ range only over operators respecting the symmetry. 
Based on intuition from statistical mechanics, stable clustering is expected to hold in the interiors of a broad class of thermal phases, including those displaying discrete symmetry-breaking order\footnote{Models displaying discrete symmetry-breaking order, \ie the Ising model, satisfy our notion of clustering since all symmetric local observables have rapidly decaying correlation functions.} and quantum topological order. Violations of the condition are expected near finite- and zero-temperature phase-transition points due to the presence of long-range correlations.

A notable feature of our construction is that the channel circuit satisfies \textbf{local reversibility} (formally defined in Def.\ref{def:LR}), which intuitively means that each channel gate’s action does not disturb long-distance correlations in the state at that time, and therefore can be reversed locally. Locally reversible channel circuits were proposed in Ref.~\cite{2025localreversibility} as a generalization of local unitary circuits to open quantum systems. Among other consequences, this implies the existence of a reversal circuit that transforms the final state back to the initial state, thereby connecting to the two-way connectivity definition of mixed-state phases. It also guarantees that the circuit preserves sharp features of the state, including symmetries and their anomalies~\cite{2025localreversibility}.

As an application of the construction, we prove a connection between self-correcting quantum memory and quantum thermal phases. A Hamiltonian system is called a self-correcting quantum memory if it is capable of preserving quantum information for a macroscopically long time when coupled to a low-temperature bath. Prominent examples include toric codes in $D\geq4$ dimensions~\cite{dennis2002topological, 20084DTC}. At first sight, this property relies on both the system’s Hamiltonian and the specific form of the system–environment interaction. Therefore, it is \textit{a priori} unclear whether (a) self-correction can be inferred from the steady state being a low-temperature Gibbs state alone, and (b) if it is a universal property shared by other systems in the same thermal phase. We answer both questions affirmatively. For systems in the same thermal phase as a zero-temperature topological code, we show that any such system acts as a self-correcting quantum memory with an $\exp(L^{\mO(1)})$ memory time (Theorem \ref{thm:topo_lifetime}). As part of the proof, we construct channels that encode and decode quantum information into and from the system at finite-temperature equilibrium.

For Hamiltonians composed of commuting local terms, we can relax the stable clustering condition to a clustering condition only and obtain an improved construction: if two Gibbs states are connected by a path along which clustering (rather than stable clustering) holds, then there exists an $\mO(\polylog(\poly L/\epsilon))$-local Lindbladian evolution that approximately transforms one state into the other (Theorem \ref{thm:cont_connectivity}). We expect a similar relaxation to be possible for the noncommuting case as well and leave it for future work.

We note that our results build on several recent advances in quantum information theory. In particular, we use the result, proved in \cite{2025CR}, that the Gibbs state of a local Hamiltonian is approximately Markovian, and that, under clustering of correlations, a local perturbation of the Hamiltonian affects only a nearby region of a Gibbs state. Work on Glauber dynamics for the classical Ising model \cite{2021IsingGlauber}, although not phrased explicitly in this manner, also suggests that the symmetry-broken phase of the Ising model possesses some form of circuit connectivity. We further note that a recent work on self-correcting quantum memories studies related questions in the context of stabilizer Hamiltonians, particularly the 4D toric code~\cite{2025BGL}.

Finally, we remark that this work takes a state-centric point of view to study thermal phases. Alternatively, one can take a dynamics-centric point of view by formulating and studying phases of thermalizing dynamics, where Gibbs states emerge as steady states. As exemplified by our result on self-correcting quantum memories (Theorem~\ref{thm:lifetime}), many properties of the dynamics can already be inferred from its steady state. Nevertheless, there are also properties of the dynamics that are not reflected sharply in the state, with a notable example in finite dimensions being the domain-wall roughening transition of the 3D Ising model~\cite{hasenbusch1996roughening}. The transition, which occurs at a critical temperature within the Ising model's ordered phase, is a sharp change in the lifetime of a large, traversing domain wall. Interestingly, it is not accompanied by any singular behavior in the Gibbs state, and we expect that Gibbs states below and above the critical temperature can be connected by the channel circuits we construct. More drastic examples, including ones displaying multiple dynamical phase transitions without any thermodynamic transitions, can be found in systems on expander graphs~\cite{eggarter1974cayley,montanari2006rigorous, hong2025quantum, placke2025expansion, placke2024topological}. Characterizing these dynamical phases and transitions in a unified framework remains a challenging open problem.

\section{Setup and Definitions}

We consider a system on a $D$-dimensional regular lattice, where $\Lambda$ denotes the set of all sites. For any subset $X \subset \Lambda$, let $|X|$ denote the number of sites in $X$. For two disjoint subsets $X$ and $Y$, we define the distance $d(X,Y)$ as the length of the shortest path between them. For an operator $O$, $\opnorm{O}$ and $\opnorm{O}_1$ denote its operator norm and trace norm, respectively.

We focus on systems with local interactions in thermal equilibrium, described by the Gibbs states:
\begin{equation} 
\rho_\bbeta := \frac{e^{-H_\bbeta}}{\Tr(e^{-H_{\bbeta}})},
\quad\quad
H_{\bbeta} := \bbeta\cdot\bh =\sum_{Z} \beta_Z h_Z
\label{eq:Gibbs}
\end{equation}
where $\bh=\{h_Z\}$ is a list of local interaction terms with $\Omega(L^D)$ elements. Each $h_Z$ acts on a region $Z\subset \Lambda$ with $\operatorname{diam}(Z)\leq R$ and $\Vert h_X \Vert \leq 1$. The inverse temperature of $\rho_\bbeta$ is defined as $|\bbeta|_\infty:=\sup_Z|\beta_Z|$, in agreement with the literature (\eg \cite{2025CR}).

We henceforth assume a fixed set of interactions $\bh$, chosen to specify the family of Hamiltonians under consideration. Three choices of $\bh$ are relevant to this work: (i) all local operators with bounded support, which gives the generic local-Hamiltonian family; (ii) all local operators that are symmetric under a global onsite symmetry, which gives the family of symmetric local Hamiltonians (e.g.\ Ising-symmetric models); and (iii) a set of mutually commuting local operators, which covers all classical Hamiltonians and stabilizer Hamiltonians. 

We use $\A$ to denote the operator algebra generated by $\bh$. For a subregion $X\subseteq\Lambda$, we use $\A_X$ to denote the algebra generated by elements of $\Tr_{\bar X}\A$. We observe that $\rho_\bbeta\in\A$ and $\rho_\bbeta^X\in\A_X$ for any $\bbeta$.

We now formally define the clustering of correlations condition used in this work. For a given operator algebra $\A$ and a state $\rho\in\A$, we define the covariance of two disjoint subsets $X, Y \subset \Lambda$ as:
\begin{equation}
    \cov^{\A}_\rho(X,Y): = \sup_{\substack{O_1\in \mathcal{A}_X: \Vert O_1 \Vert = 1\\ O_2\in \mathcal{A}_Y: \Vert O_2 \Vert = 1}} |\braket{O_1 O_2}_\rho - \braket{O_1}_\rho\braket{O_2}_\rho|,
    \label{eq:covariance}
\end{equation}
where $\braket{O}_\rho=\Tr(O \rho)$. {
\begin{definition}[Clustering of correlations]\label{def:clustering}
    Let $\xi\geq 0$ be a constant and $f(x)$ be a function that grows at most polynomially with $x$. We say a state $\rho\in\A$ exhibits $(\xi, f)$ clustering if, for every annulus tripartition in Fig.~\ref{fig:annulus}, the following holds:
\begin{equation}
    \cov^{\A}_\rho (A,C) \leq f(L) e^{-d(A,C)/\xi},
    \label{eq:decayofc}
\end{equation}
\end{definition}
}
We remark that this clustering condition is weaker than more common versions in the literature (\eg\cite{2025LPPL}) due to the restriction to operators from $\A_A$ and $\A_C$, as well as the particular choice of the A-B-C geometry illustrated in Fig.\ref{fig:annulus}. {This makes it applicable to a large class of states with nontrivial orders. For instance, the Ising model below the critical temperature satisfies our definition of clustering, since local operators in $\A_{A/C}$ respect the Ising symmetry and display exponentially decaying correlations, even though nonsymmetric ones have long-range correlations. As another example, topologically ordered ground states on a torus satisfy our notion of clustering, even though they have long-range correlations for non-annulus partitions (see, e.g., \cite{konig2014generating}, which shows that noncontractible loop operators display long-range correlations).}

For Gibbs states, we introduce another concept dubbed \emph{stable clustering}: 
\begin{definition}[Stable clustering]\label{def:stable_clustering}
A Gibbs state $\rho_\bbeta$ (as defined in Eq.\eqref{eq:Gibbs}) satisfies $(\delta, \xi, f)$ stable clustering if, for any $\Delta\bbeta$ with $|\Delta\bbeta|_\infty\leq \delta$, the Gibbs state $\rho_{\bbeta+\Delta\bbeta}$ exhibits $(\xi, f)$ clustering. 
\end{definition}

Henceforth, when we use the term `stable clustering' without specifying $(\delta, \xi, f)$, it is assumed that $\delta,\xi=\mO(1)$ and $f=\poly L$.


We expect stable clustering to be satisfied in the interior of certain classes of thermal phases. Here, `certain classes' include topologically ordered phases, spontaneous symmetry-breaking (SSB) phases of discrete global symmetries (if $\bh$ includes only symmetric terms), and the high-temperature trivial phase.
Exceptions should include phases displaying SSB of continuous global symmetries or spatial symmetries; for the former, due to the presence of Goldstone modes leading to power-law-decaying correlations, and for the latter, since the perturbation $\Delta\bbeta$ can explicitly break spatial symmetry. {Stable clustering is also expected to be violated at phase transitions, including both discontinuous (first order) and continuous (higher order) ones.}

{
In open systems, local channel circuits are natural generalizations of unitary circuits. On a lattice $\Lambda$, a local channel circuit of depth $T$ is a quantum channel of the form:
\begin{equation}
    \C = \C_T \circ \cdots \circ\C_2\circ\C_1
    \quad{\rm where}\quad
    \C_t = {\prod}_{x}\E_{t,x}
\end{equation}
where each $\C_t$ is called a layer of the circuit, and each $\E_{t,x}$ is called a gate in layer $t$. Different gates within a layer have non-overlapping spatial support. The range of $\C$, denoted by $\rng\C$, is defined as $T$ times the largest diameter of the supports of the $\E_{t,x}$.

The channel circuits we construct in later sections satisfy a strong property called local reversibility~\cite{2025localreversibility}, formally defined as:
\begin{definition}[Local reversibility (LR)]
\label{def:LR}
Let $\C$ be a channel circuit as defined above. $\C$ is $\epsilon_{LR}$-locally reversible with respect to an input state $\rho$ if, for each channel gate $\E_{t,x}$, there exists another channel $\tilde\E_{t,x}$ with the same spatial support as $\E_{t,x}$ such that
\begin{equation}\label{eq:LR}
        \opnorm{\tilde\E_{t,x}\circ\E_{t,x}\circ\C'[\rho] - \C'[\rho]}_1\leq \epsilon_{LR}
\end{equation}
where $\C'=\C_{t-1}\cdots\C_1$ represents the first $(t-1)$ layers of $\C$.
\end{definition}
Intuitively, local reversibility means each channel gate’s action does not disturb long-range correlations of the state at that time and can therefore be reversed locally. This gate-by-gate reversibility, among other consequences, implies the existence of a reversal circuit:
\begin{equation}\label{eq:rev_circuit}
    \tilde\C= \tilde\C_1 \circ \cdots \circ\tilde\C_{T-1}\circ\tilde\C_T
    \quad{\rm where}\quad
    \tilde\C_t = {\prod}_{x}\tilde\E_{t,x},
\end{equation}
that reverses $\C$’s action on $\rho$:
\begin{equation}
    \opnorm{\tilde\C\circ\C[\rho]-\rho}_1 \leq |\C|\cdot\epsilon_{LR}.
    \label{eq:rev_c_err}
\end{equation}
Here $|\C|$ denotes the total number of gates within $\C$. A proof of Eq.\eqref{eq:rev_c_err} can be found in App.\ref{app:LR_LI}. 
}

\section{Constructing channel circuits connecting Gibbs states}\label{sec:circuit}

We now state our first result:
\begin{theorem}
    \label{thm:connectivity}
Let $\rho_{\bbeta}$ and $\rho_{\bbeta'}$ be two Gibbs states connected by a path in $\bbeta$-space along which stable clustering is satisfied. Then there exists a $2\epsilon_\C$-locally reversible channel circuit $\mathcal{C}$ with range $\rng\C = \polylog(\poly(L)/\epsilon_\C)$, such that $\opnorm{\rho_{\bbeta'} - \C[\rho_{\bbeta}]}_1\leq\epsilon_\C$. Here $L$ is the linear system size.
\end{theorem}
We remark that our construction and the error bound continue to work when one of (or both of) $\rho_{\bbeta}$ and $\rho_{\bbeta'}$ is (are) zero-temperature gapped ground state(s), if we impose some sparsity conditions on the density of states near ground states, similar to the condition introduced in~\cite{hastings2007quantum}. We discuss this point in detail in Sec.\ref{sec:zeroT}.

Before proving the theorem in the next two subsections, we review several results on which our proof is based.

The first result relates clustering to the property of \textit{local perturbation perturbs locally} (LPPL). As its name suggests, LPPL concerns the extent to which a local change in the Hamiltonian modifies {the expectation value of } distant observables 
{evaluated in}
the Gibbs state. The theorem is formally stated below. The proof of this result, which closely follows the proof of Theorem 25 in Ref.~\cite{2025LPPL}, is included in App.\ref{app:LPPL}.

\begin{lemma}[Clustering implies LPPL. Theorem 25 in Ref.~\cite{2025LPPL}, modified]
Let $\rho(s)$ be the Gibbs state corresponding to Hamiltonian $H(s) = H(0)+s\cdot h_A$, with $h_A$ supported on a local region $A$. Then, for any region $C$ non-overlapping with $A$ and any $l<d(A,C)$, one has:
\begin{equation}
\begin{split}
    &\opnorm{\rho^C(0)-\rho^C(1)}_1 \\
    \leq 
    & \opnorm{h_A}\cdot\left(\sup_{s \in [0, 1]} \cov^\A_{\rho(s)}(A_{+l}; C) +  6|A| e^{-\frac{1}{1+v/\pi}l}\right).
 \end{split}
 \label{eq:clustering2lppl}
\end{equation}
where $A_{+l}$ denotes the set of sites whose distance to $A$ is no larger than $l$, and $v$ is the maximum Lieb-Robinson velocity of $\{H(s)\}_{s\in[0,1]}$. 
\end{lemma}

The second result states that Gibbs states of local Hamiltonians satisfy so-called approximate Markov property. The property is defined via quantum conditional mutual information (CMI). For any state defined on a tripartite system $ABC$, CMI is defined as:
\begin{equation}
    I(A:C|B)_\rho := I(A:BC)_\rho - I(A:B)_\rho
\end{equation}
where $I(X:Y)_\rho$ is the quantum mutual information, a measure of correlations between two parties. For Gibbs states on finite-dimensional lattices, we consider the annulus tripartition (Fig.~\ref{fig:annulus}): $A$ is a small simply connected region, $B$ is a buffer region surrounding $A$, and $C$ is the rest of the system. In this context, CMI measures the locality of correlations: small $I(A:C|B)$ means that $A$’s correlations with the rest of the system can be well captured by the buffer $B$—a property called the Markov property. We remark that clustering and the Markov property are not directly related: there are systems displaying clustering but not the Markov property, and vice versa. The following theorem states that all Gibbs states satisfy an approximate Markov property locally:
\begin{lemma}[Gibbs states are locally Markovian. Ref.\cite{2025CR}]
    For a Gibbs state $\rho_\bbeta$ and any annular-shaped tripartition depicted in Fig.~\ref{fig:annulus}, the CMI is bounded as:
\begin{equation} 
\begin{split}
    I(A:C|B)_{\rho_\bbeta}  \leq  |A|\cdot|C| e^{\kappa + \mu|A| -  d(A,C)/\lambda}, 
\end{split}
\label{eq:CMI}
\end{equation}
where $\kappa$, $\mu$, and $\lambda$ are system-size-independent constants growing at most polynomially with $|\bbeta|_\infty$.
\end{lemma}
We note that for any set of $\bbeta$ with $\max|\bbeta|_\infty$ bounded above, one can choose $\kappa, \mu, \lambda=\poly(\max|\bbeta|_\infty)$ such that Eq.~\eqref{eq:CMI} holds uniformly for all $\bbeta$ in the set. For commuting local Hamiltonians a stronger claim holds: 
\begin{equation}
     I(A:C|B)_{\rho_\bbeta} = 0\quad{\rm when}\quad {\rm dist}(A,C)>R
\end{equation}
which is the content of the quantum Hammersley–Clifford theorem~\cite{2008LeiferPoulin,2012commuting}.

The importance of the Markov property to our construction lies in its connection to recoverability. This is stated precisely as follows:
\begin{lemma}[Markov property implies recoverability. Ref.~\cite{2015JRWW}, modified]
For a state $\rho$ defined on a tripartite system $A\cup B\cup C$, there exists a quantum channel $\R^\rho_{B\to AB}$ acting from $B$ to $A\cup B$ such that:
\begin{equation}
        \opnorm{\rho - \R^\rho_{B\to AB}\circ\Tr_A(\rho)}_1\leq\sqrt{4\ln 2 \cdot I(A:C|B)_\rho} .
        \label{eq:approxMarkov}
    \end{equation}
\end{lemma}
An explicit construction of $\R$ is given by the twirled Petz recovery map (see App.~\ref{app:recovery} for an explicit expression and a proof of the lemma using the main theorem of Ref.~\cite{2015JRWW}).

\subsection{Local Interaction Variation}\label{subsec:local_variation}
\begin{figure}
\begin{center}
  \includegraphics[width=.25\textwidth]{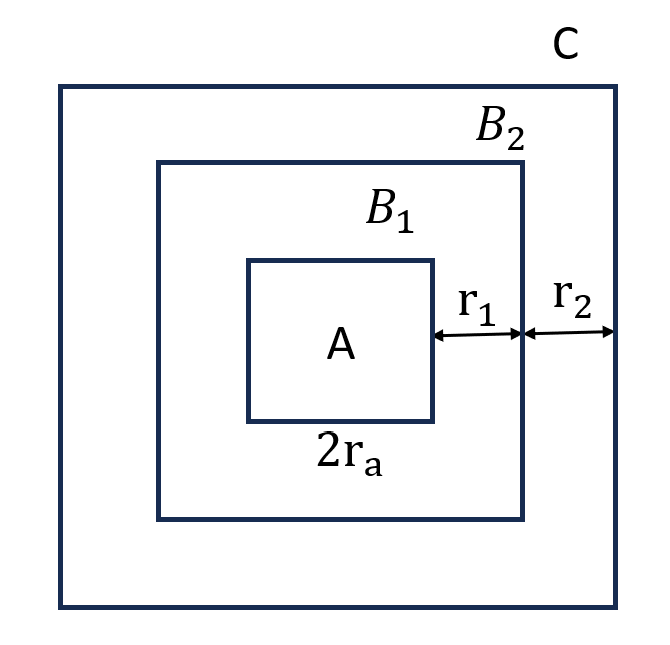} 
\end{center}
\caption{Geometry for local interaction variation.}
\label{fig:temperaturevariation}
\end{figure}

{We start by constructing a channel that transforms between two Gibbs states whose Hamiltonians differ only in a local region $A$ [see Fig.\ref{fig:temperaturevariation}].  Let $\bbeta$ satisfy $(\delta, \xi, f)$-stable clustering. We aim to find a channel that changes $\rho = \rho_{\bbeta}$ into $\tilde\rho = \rho_{\bbeta+\Delta\bbeta_A}$ for some simply connected region $A$ and an interaction variation $\Delta\bbeta_A$ such that $\Delta\bbeta_{A}\cdot\bh$ is supported strictly within $A$ and $|\Delta\bbeta_A|_\infty<\delta$.}

Let $B=B_1\cup B_2$ be an annulus-shaped region surrounding $A$, and $C$ be the rest of the system [Fig. \ref{fig:temperaturevariation}]. We use $r_a, r_1, r_2$ to denote the radii of $A, B_1, B_2$, respectively. 

We consider the following channel acting on $AB$:
\begin{equation}
    \M= \R^{\tilde\rho}_{B_2\rightarrow A B}\circ\Tr_{AB_1}
    \label{eq:localM}
\end{equation}
where $\R_{B_2\rightarrow AB}^{\tilde\rho}$ is the twirled Petz recovery map using $\tilde\rho$ as a reference state (an explicit form of $\R$ can be found in App.\ref{app:recovery}). The intuition behind Eq.\eqref{eq:localM} is to first erase $AB_1$ from $\rho$; by LPPL, the remaining state on $B_2C$ is similar to $\tilde \rho^{B_2C}$. One then locally resamples the state within $AB_1$ according to $\tilde \rho$ using the recovery map in Eq.\eqref{eq:approxMarkov}.

As we show, $\M$ approximately transforms $\rho$ into $\tilde{\rho}$, with an error bounded by:
\begin{equation}
\begin{split}
&\opnorm{\tilde{\rho} - \M(\rho)}_1\leq\epsilon_\M\\
& \epsilon_\M = \poly(L) (e^{\kappa + \mu |A B_1| - r_2/\lambda} +e^{-r_1/\xi'} ),  
\end{split}
\label{eq:localerror}
\end{equation}
where $\xi' = \xi + 1+ v/\pi$, with $v$ being the maximum Lieb–Robinson velocity~\cite{2025LPPL} of the Hamiltonians in $\{H_{\bbeta + s\Delta\bbeta_A}\}_{s\in[0,1]}$. The other constants $\kappa$, $\mu$, and $\lambda$ are as in Eq.\eqref{eq:CMI}.

The error bound Eq.~\eqref{eq:localerror} is derived by invoking Lemmas 1–3. Using the triangle inequality (sub- and superscripts of the recovery map $\R$ are omitted):
\begin{equation}
\begin{split}
    &\opnorm{\tilde{\rho} - \M[\rho]}_1\\ 
    \leq  &\opnorm{\tilde{\rho}-\R[\tilde\rho^{B_2C}]}_1 + \opnorm{\R[\tilde\rho^{B_2 C}]-\R[\rho^{B_2 C}]}_1.
\end{split}
\end{equation}
The approximate Markov property of Gibbs states (Lemmas 2 and 3) guarantees that the first term is bounded by the $e^{\kappa + \mu |AB_1| - r_2/\lambda}$ term. Further, since $\rho$ satisfies stable clustering, Lemma 1 implies $\rho$ and $\tilde\rho$ are approximately indistinguishable on $B_2C$:
\begin{equation}
\begin{split}
    &\opnorm{\tilde{\rho}^{B_2C} - \rho^{B_2 C}}_1 \\
    \leq~ &\delta\cdot(f(L)e^{-\frac{1}{\xi}(r_1-l)}+6|A| e^{-\frac{1}{1+v/\pi}l})\quad \forall l<r_1\\
    =~ &\poly(L) e^{-r_1/\xi'},
\end{split}
\end{equation}
where the last equality is obtained by letting $l=\left(1+\frac{\xi}{1+v/\pi}\right)^{-1} r_1$, which makes the two exponential factors equal. The corresponding $\xi'$ is $\xi' = \xi+1+v/\pi$. This finishes the proof of Eq.\eqref{eq:localerror}.

We now consider the trade-off between $\M$'s locality and the approximation error. To ensure $\opnorm{\M[\rho]-\tilde\rho}_1\leq\epsilon_\M$, it suffices to choose
\begin{equation}
    \begin{split}
        & r_1=r_a\geq \xi' \ln({\poly(L)}/{\epsilon_\M})\\
        & r_2\geq \kappa\lambda+\mu \lambda (4 r_1)^D+\lambda \ln({\poly(L)}/{\epsilon_\M}).
    \end{split}
\end{equation}
If we further assume the inverse temperature $|\bbeta|_\infty$ is at most $\log(\poly(L)/\epsilon_\M)$, we conclude that
\begin{equation}
    r_a, r_1, r_2 = \polylog(\poly L/\epsilon_\M).
    \label{eq:tradeoff}
\end{equation}

Before proceeding, we note that the channel $\M$'s action on $\rho$ can be approximately reversed by 
\begin{equation}\label{eq:tildeM}
    \tilde\M:= \R^{\rho}_{B_2\rightarrow AB}\circ\Tr_{AB_1},
\end{equation}
which is of the same form as $\M$ but replaces $\tilde\rho$ by $\rho$ in the recovery map $\R$. Following the same proof, we have $\opnorm{\tilde\M[\tilde\rho] - \rho}_1\leq \epsilon_\M$, and therefore 
\begin{equation}
    \opnorm{\tilde \M\circ\M[\rho] - \rho}_1\leq 2\epsilon_\M. 
\end{equation}

\subsection{Global Interaction Variation}
{
Building on local interaction variation, we now describe a layered channel circuit that maps between two Gibbs states $\rho_{\bbeta}$ and $\rho_{\bbeta'}$ connected by a path satisfying stable clustering. The strategy is to make small, local changes to the Hamiltonian each time, in a carefully chosen order, so that the overall circuit implementing the global interaction variation has bounded range.}

We first observe that it suffices to construct a circuit $\G$ that maps $\rho_{\bbeta}$ to $\rho_{\bbeta+\Delta\bbeta}$ with $|\Delta\bbeta|_\infty\leq\delta$ ($\delta$ as in the stable clustering definition Def.\ref{def:stable_clustering}), and then repeat the procedure $n$ times. This is because we can discretize the path connecting $\bbeta$ to $\bbeta'$ into a sequence of length $n\sim|\bbeta-\bbeta'|_{\infty}/\delta$ ($\delta$ as in Def.\ref{def:stable_clustering}) $\bbeta_0$-$\bbeta_1$-...-$\bbeta_n$, such that $\beta_0=\bbeta, \bbeta_n=\bbeta'$ and $|\bbeta_m-\bbeta_{m-1}|_\infty\leq\delta$ ($0 < m \le n$).

\begin{figure}
\begin{center}
\includegraphics[width=.23\textwidth]{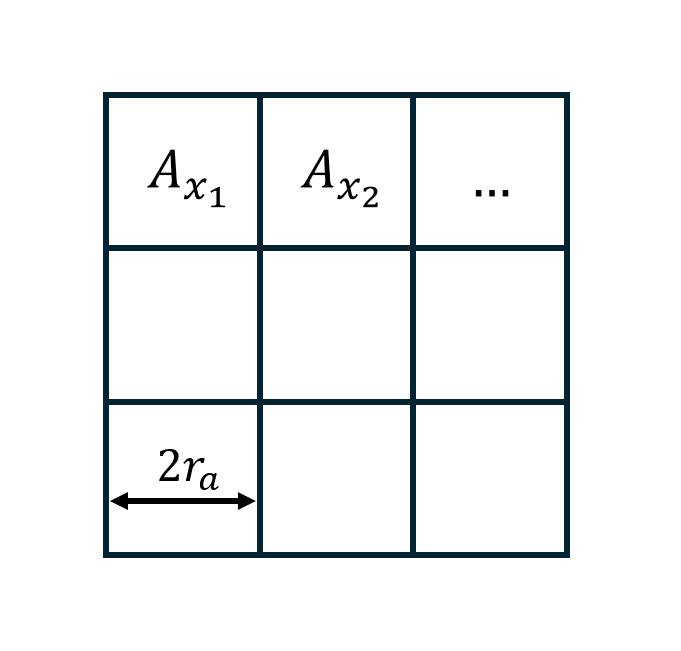}
\includegraphics[width=.23\textwidth]{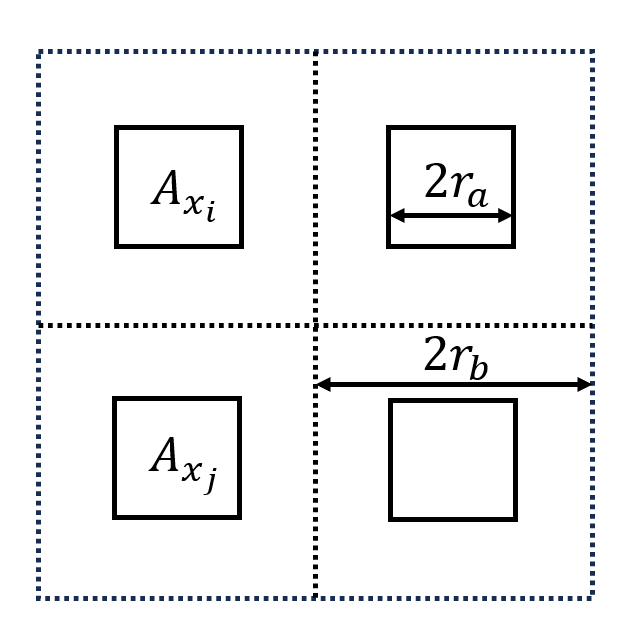} 
\end{center}
\caption{Illustration of global interaction variation. 
(left) The system is partitioned into a set of diameter $2r_a$ blocks $\{A_{x_i}\}_i$. Adjacent blocks have a width $R$ overlap on their boundary, which is omitted in the figure. (right) The gate $\M^{(i)}$ used for varying interaction on $A_{x_i}$ supports within the dashed box containing $A_{x_i}$. Two gates $\M^{(i)}$ and $\M^{(j)}$  can be placed in a single layer and implemented in parallel if $x_i$ and $x_j$ are at least $2r_b$ apart.
}
\label{fig:global}
\end{figure}

To this end, we begin by partitioning the system into hypercubic blocks in $D$ spatial dimensions, 
with each block having diameter $2 r_a$. Two adjacent blocks overlap along their boundaries in a region of width $R$ (recall that $R$ is the range of interaction terms in $\bh$). We denote each block region by $A_x$, with $x$ being the center of the block. There are $N = L^D/(2r_a-R)^D$ blocks in total. 

This accordingly leads to a partition of the Hamiltonian terms: $\bbeta\cdot\bh = \sum_{x} \bbeta_x\cdot\bh_x$, where $\bh_x$ contains only $h_Z$'s supported entirely within $A_x$ and $\bbeta_x$ are their coefficients. Due to the width-$R$ overlap, each $h_Z$ is contained in at least one $A_x$. If an $h_Z$ is in more than one block, it is assigned to only one of them.

We observe that for any ordering of blocks $\{A_{x_1}, A_{x_2},...,A_{x_N}\}$ we can turn $\rho_\bbeta$ into $\rho_{\bbeta+\Delta\bbeta}$ by varying interactions on different blocks sequentially, using the scheme described in the previous section. More specifically, for $i=0,1,\ldots,N$ let
\begin{equation}
    \rho^{(i)} := \rho_{\bbeta+\Delta\bbeta_i}\quad{\rm where}\quad
    \Delta\bbeta_i \cdot\bh :={\sum}_{j\leq i}\Delta\bbeta_{x_j}\cdot\bh_{x_j},
    \label{eq:intermediatestates}
\end{equation}
and let $\M^{(i)}$ be the channel defined in Eq.\eqref{eq:localM} with $\tilde\rho$ set to $\rho^{(i+1)}$. Then one can show that the error of the global interaction variation is bounded by the sum of the errors in the local steps (see App.\ref{app:totalerror} for a detailed derivation):
\begin{equation}
    \begin{split}
        &\epsilon_\G := \opnorm{\G[\rho_{\bbeta}]-\rho_{\bbeta+\Delta\bbeta}}_1\\
        &\epsilon_\G\leq \sum_{i=0}^{N-1}\opnorm{\M^{(i)}[\rho^{(i)}]-\rho^{(i+1)}}_1
        \leq
        N\cdot\epsilon_{\M}.
    \end{split}
    \label{eq:totalerror}
\end{equation}
where $\G:=\M^{(N)}\circ\cdots\circ\M^{(1)}$.

We devise an optimal ordering of blocks so that the resulting $\G$ has minimal depth and range. We note that $\M^{(i)}$ and $\M^{(j)}$ are non-overlapping and can be implemented in parallel if $x_i$ and $x_j$ are at least $2r_b :=2(r_a+r_1+r_2)$ apart. Therefore, a single layer can accommodate $(L/r_b)^D$ non-overlapping gates by placing gates on an $r_b$-spaced grid (see Fig.\ref{fig:global}). In this way, $\G$ can be organized into a circuit with at most $N\cdot (L/r_b)^{-D}\approx(r_b/r_a)^D$ layers.

The total circuit $\C=\circ_{m=1}^n\G_m$, which transforms $\rho_{\bbeta}$ to $\rho_{\bbeta'}$ along the path $\rho_{\bbeta_1}$,...,$\rho_{\bbeta_n}$, has $n$ times the depth and approximation error of each $\G_m$. Each $\G_m$ transforms $\rho_{\bbeta_{m-1}}$ to $\rho_{\bbeta_{m}}$ following the construction above. Combining Eq.\eqref{eq:totalerror} and Eq.\eqref{eq:tradeoff}, we conclude that having
\begin{equation}
\rng\C = \polylog(\poly L/\epsilon_\C)
\end{equation}
is sufficient to ensure $\opnorm{\C[\rho_{\bbeta}]-\rho_{\bbeta'}}_1 \leq \epsilon_\C$, provided the inverse temperatures are bounded above by $|\bbeta|_\infty,|\bbeta'|_\infty\leq\log(\poly L/\epsilon_\C)$.  In addition, as we prove in App.\ref{app:LR_LI}, the constructed $\C$ is $2\epsilon_{\C}$-LR with respect to $\rho_{\bbeta}$. This completes the proof of Theorem \ref{thm:connectivity}.

\subsection{Zero-temperature limit}\label{sec:zeroT}

In this section we discuss how our construction can be applied when the initial (or equivalently the target) Gibbs state is a maximally mixed state in the ground-state subspace of some $H_{\bbeta}$\footnote{Following standard definition, ground-state subspace includes all states whose energy splitting from the ground state is smaller than any $\poly^{-1}L$.}, denoted by $\rho_{\rm g.s.}$. We assume $|\bbeta|_\infty=1$ in this section without loss of generality.

When there is an energy gap above the ground-state subspace, $\rho_{\rm g.s.}$ can be well approximated by $\rho_{s\bbeta}$ for large enough $s$ (low enough temperature) in many physical Hamiltonians.
Following the argument in Ref.\cite{2007Has}, if we assume that low-lying excitations of $H_\bbeta$ are all closed $p$-dimensional manifolds (\eg particles, loops, membranes, etc), then the number of low-lying states grows at most polynomially with the system size, leading to the approximation error (see App.\ref{app:groundstates} for a detailed argument and derivations):
\begin{equation}
    \opnorm{\rho_{\rm g.s.} -\rho_{s\bbeta}}_1=\poly(L)e^{-s\Delta}
    \label{eq:dos_bound}
\end{equation}
where $\Delta=\mO(1)$ is $H_\bbeta$'s energy gap above the ground-state subspace.

Hence, to guarantee an $\epsilon$ approximation error, we need $s$ to be as large as $\log(\poly L/\epsilon)$. Now we can follow the methods in previous sections and construct a channel circuit $\C$ of range $\polylog(\poly L / \epsilon)$ that transforms $\rho_{s\bbeta}$ to the final state $\rho_{\bbeta'}$, with error also bounded by $\epsilon$. This leads to an error bound:
\begin{equation}
\begin{split}
    &\opnorm{\C[\rho_{\rm g.s.}]-\rho_{\bbeta'}}_1\\
    \leq
    &\opnorm{\C[\rho_{s\bbeta}]-\rho_{\bbeta'}}_1 + \opnorm{\rho_{s\bbeta}-\rho_{\rm g.s.}}_1\leq 2\epsilon.
\end{split}
\end{equation}
Following an analogous procedure, we can handle the case in which $\rho_{\bbeta'}$ approaches the zero-temperature limit.

\section{Thermal phases of self-correcting quantum memories}

As an application of Theorem \ref{thm:connectivity}, in this section we show that quantum self-correction is a universal property shared by all states within the same thermal phase as a zero-temperature topological quantum code. {More specifically, if $\C$ is the constructed circuit (following the recipe in the previous section) connecting a quantum code to another Gibbs state $\rho_\bbeta$, we prove that $\C$ encodes a pure codeword state into a mixed state that is locally indistinguishable (in the sense of Def.\ref{def:LI}) but globally distinguishable from $\rho_\bbeta$. Due to the local indistinguishability from $\rho_\bbeta$, the encoded state is long-lived under any quasi-local dynamics with $\rho_\bbeta$ as a steady state.}

We start by formulating the self-correcting property formally. We model the environment’s influence on the system, as well as the system’s internal evolution, jointly by a quasi-local Lindbladian super-operator of the form:
\begin{equation}
    \mL = {\sum}_X \mL_X
\end{equation}
such that each $\mL_X$ satisfies $\Vert \mL_X \Vert_{1\to1}\leq 1$. Quasi-locality of each $\mL_X$ means there exists an approximating family of diameter-$l$-support super-operators $\{\mL_X^l \}$, such that $\Vert \mL_X - \mL_X^l \Vert_{1\to1} \leq e^{-\alpha l}$ with a constant $\alpha>0$. 
\footnote{We consider quasi-local rather than strictly local Lindbladians since, for any Gibbs state $\rho_{\bbeta}$ of local Hamiltonians, there exists a quasi-local Lindbladian $\mL=\sum_X \mL_X$ having $\rho_\bbeta$ as a frustration-free steady state, \ie $\mL_X[\rho_{\bbeta}]=0$ for each $X$~\cite{2023Gibbssampler}. It is unclear whether strictly local Lindbladians with the same property exist for generic noncommuting Hamiltonians.}
Since the system’s Gibbs state $\rho_\bbeta$ emerges as the steady state of system–environment interaction, it follows that $\mL[\rho_\bbeta]=0$. Here we assume the stronger property $\mL_X[\rho_\bbeta]=0$, which means $\rho_\bbeta$ is a frustration-free steady state of $\mL$. Concrete instances of $\mL$ satisfying all properties above have recently been constructed in \cite{2023Gibbssampler}.

We say a Lindbladian $\mL$ preserves $\log_2 K$ qubits of quantum information for a memory time $t$ with error $\epsilon$ if there exists an encoding channel ${\rm En}:\S(\mathbb{C}^{K})\mapsto\S(\H)$ (here $\H$ denotes the system's Hilbert space, and $\S(x)$ denotes the set of density operators defined on a Hilbert space $x$) and a decoding channel ${\rm De}:\S(\H)\mapsto\S(\mathbb{C}^{K})$ such that:
\begin{equation}
    \opnorm{{\rm De}\circ e^{\tau\mL}\circ {\rm En}(\omega)-\omega}_1 \leq \epsilon 
    \quad \forall \omega\in\S(\mathbb{C}^{K}),\ \tau\leq t.\label{eq:def_memtime}
\end{equation}

{We call a local Hamiltonian's zero-temperature Gibbs state $\Pi$ (which, as we recall, is the maximally mixed state in its ground-state subspace $V$) a diameter-$\ell$ quantum code if $\Pi O\Pi\propto\Pi$ for any operator $O$ supported on a region with diameter no larger than $\ell$ \footnote{We emphasize that a code's diameter is not its distance: a quantum code has distance $d$ if $\Pi O\Pi\propto\Pi$ for any operator $O$ supported on a region with \emph{number of sites} no more than $d$.}. } Topological codes, \ie quantum codes whose code states display quantum topological order, have diameter at least $L/2$.

{As a crucial ingredient of the proof that self-correction is a universal property, the following theorem shows that whether a local dynamics can preserve quantum information may be inferred from the properties of its steady state.}

\begin{theorem}\label{thm:lifetime}
Suppose a state $\rho$ satisfies $\opnorm{\C(\Pi)-\rho}_1\leq\epsilon_\C$ for a diameter-$\ell$ code $\Pi$ and a channel circuit $\C$. Further, suppose $\C$ is $\epsilon_{LR}$-locally reversible with respect to $\Pi$ and satisfies $\rng\C\leq \ell/4$. Then any quasi-local Lindbladian $\mL$ having $\rho$ as a frustration-free steady state preserves $\log_2({\rm rank}\,\Pi)$ qubits of information for time $t$ with error
\begin{equation}
    \epsilon_t = \poly(L)(\epsilon_{LR}+\epsilon_{\C}+e^{-\alpha \ell/2})\cdot (t+1).
\end{equation}
\end{theorem}
We remark that we do not assume $\mL$ to satisfy quantum detailed balance, and therefore the theorem is applicable to Lindbladian dynamics beyond thermalization. 

The proof of Thm.\ref{thm:lifetime} utilizes \textit{local indistinguishability} (LI) of states within a topological code space, formally defined below. 
\begin{definition}[Local indistinguishability (LI) for mixed states~\cite{2025localreversibility}]
\label{def:LI}
    Two states $\rho$ and $\sigma$ are $(r, \epsilon_{LI})$-locally indistinguishable if, for any simply connected $A$ with $\diam(A)\leq r$, there exist channels $\D, \D'$ acting outside $A$ such that $\opnorm{\D[\rho]-\sigma}_1\leq\epsilon_{LI}$ and $\opnorm{\D'[\sigma]-\rho}_1\leq\epsilon_{LI}$. 
\end{definition}

It is straightforward to see that if $\rho$ and $\sigma$ are LI, then $\rho^A\approx\sigma^{A}$ for any $A$ with $\diam(A)\leq r$ since the channel $\D$ (or $\D'$) does not act on $A$, hence the name ``local indistinguishability'' 
\footnote{The converse does not hold in general: two states can agree on all local observables without being related by a channel acting outside $A$. Actually, as shown in Ref.~\cite{2025localreversibility}, matching of local observables, combined with the approximate Markov property, is equivalent to Def.~\ref{def:LI}.}.
If $V$ is a code space for a diameter-$\ell$ quantum code, then any two codeword states are $(\ell, 0)$-LI (see App.~\ref{app:clear_lemma} for a proof), a conclusion known as the cleaning lemma~\cite{kalachev2022linear} in quantum coding theory.

The following lemma (whose proof can be found in App.~\ref{app:LR_LI}) shows that local indistinguishability is preserved under a locally reversible circuit.

\begin{lemma}\label{lemma:LI_LR}
    Suppose $\rho$ and $\sigma$ are $(\ell, 0)$-LI and the channel circuit $\C$ is $\epsilon_{LR}$-LR with respect to $\rho$. Then:
    \begin{enumerate}[leftmargin=0.6cm]
        \item $\C$ is $(2 |\C| \epsilon_{LR}+\epsilon_{LR})$-LR with respect to $\sigma$ with the same set of reversal gates, and
        \item $\C[\rho]$ and $\C[\sigma]$ are $(\ell-2\rng\C, (2|\C|^2+|\C|)\epsilon_{LR})$-LI.
    \end{enumerate}
    where $|\C|$ denotes the number of gates in $\C$.
\end{lemma}

We are now ready to prove Thm.~\ref{thm:lifetime}.

Following the definition in Eq.\eqref{eq:def_memtime}, we identify the subspace spanned by $\Pi$ as $\mathbb{C}^{K}$ and take the encoder and decoder to be ${\rm En}=\C$ and ${\rm De}=\tilde\C$ (recall that $\tilde\C$ is the reversal circuit of $\C$, defined in Eq.\eqref{eq:rev_circuit}). {The amount of logical error at time $t$ is upper-bounded by:
\begin{equation}
\begin{split}
    \epsilon_t := 
    &\opnorm{{\rm De}\circ e^{t\mL}\circ {\rm En}[\omega]-\omega}_1\\
    =
    &\opnorm{\tilde\C e^{t\mL} \C[\omega]-\omega}_1\\
    \leq
    & \opnorm{\tilde\C\C[\omega]-\omega}_1+
    \opnorm{\tilde\C e^{t\mL} \C[\omega]- \tilde\C\C[\omega]}_1\\
    \leq
    &\opnorm{\tilde\C\C[\omega]-\omega}_1 +\opnorm{e^{t\mL} \C[\omega]- \C[\omega]}_1
\end{split}
\end{equation}
It suffices to bound the two terms on the last line separately.}

{The first term is bounded since $\C$ is locally reversible with respect to $\omega$, as shown below.} Since $\omega$ and $\Pi$ are states within the quantum code space, they are $(\ell, 0)$-LI. According to Lemma~\ref{lemma:LI_LR}, $\C$ is $(2|\C|+1)\epsilon_{LR}$-LR with respect to $\omega$, which implies that the first term is bounded by $(2|\C|^2+|\C|)\epsilon_{LR}=\poly(L)\cdot\epsilon_{LR}$.

{Now we show that $\C[\omega]$ is an approximate frustration-free steady state of the Lindbladian $\mL$, leading to a bound on the second term. The underlying intuition is simple: since $\C$ preserves local indistinguishability, $\C[\omega]$ and $\C[\Pi]$ remain LI. Therefore, any Lindbladian term $\mL_X$ having the latter as a steady state must also leave the former almost unchanged.

We now turn this intuition into a proof with an explicit error bound.} By Lemma~\ref{lemma:LI_LR}, $\C[\omega]$ and $\C[\Pi]$ are $(\ell/2, (2|\C|^2+|\C|)\epsilon_{LR})$-LI, which by definition implies there exists a channel $\D$, which can be chosen to be supported outside any diameter-$\ell/2$ region, such that $\opnorm{\D\circ\C[\Pi]-\C[\omega]}_1\leq (2|\C|^2+|\C|)\epsilon_{LR}$. We therefore obtain the following series of approximations (the notation $A\mysim{\epsilon}B$ means $\opnorm{A-B}_1\leq \epsilon\cdot\poly L$):
\begin{equation}
\begin{split}
    &\mL_X\circ\C[\omega]
    \mysim{\epsilon_{LR}}
    \mL_X\circ\D\circ\C[\Pi]
    \mysim{\epsilon_{\C}}
    \mL_X\circ\D[\rho]
    \mysim{e^{-\alpha l}}\\
    &\mL^l_X\circ\D[\rho]
    =
    \D\circ\mL^l_X[\rho]
    \mysim{e^{-\alpha l}}
    \D\circ\mL_X[\rho] = 0
\end{split}
\end{equation}
{In the equality step, we use the fact that we can choose $\D$ so that it is non-overlapping with $\mL_X^l$, as long as $l\leq \ell/2$.}

Therefore
$\opnorm{\mL\circ\C[\omega]}_1\leq\sum_X \opnorm{\mL_X\circ\C[\omega]}_1\leq\poly(L)(\epsilon_{LR}+\epsilon_\C+e^{-\alpha l})$, which leads to
\begin{equation}\label{eq:mem_bound}
\begin{split}
     \opnorm{ e^{t\mL} \C[\omega]- \C[\omega]}_1
    &\leq
    t\cdot\opnorm{\mL\circ\C[\omega]}_1\\
    &\leq 
    t\cdot\poly(L)(\epsilon_{LR}+\epsilon_\C+e^{-\alpha l}).
\end{split}
\end{equation}
Here the first inequality uses the fact that $\opnorm{e^{t\mL}[\rho]-\rho}_1\leq t\opnorm{\mL[\rho]}_1$, which we derive in App.\ref{app:mem_bound}. Setting $l=\ell/2$ completes the proof of Thm.\ref{thm:lifetime}.

\begin{theorem}[Self-correction is a property of thermal phases]\label{thm:topo_lifetime}
Suppose $\rho_\bbeta$ is connected to a zero-temperature diameter-$L/2$ quantum code $\Pi$ by a path along which stable clustering is satisfied. Then any quasi-local Lindbladian $\mL$ having $\rho_\bbeta$ as a frustration-free steady state preserves $\log_2 ({\rm rank} \,\Pi)$ qubits of quantum information for time $t$ with error:
\begin{equation}
        \epsilon_t \leq (t+1)\cdot\exp(-L^{\mO(1)}) 
\end{equation}
for sufficiently large $L$. 
\end{theorem}
This theorem is obtained by combining Thms.\ \ref{thm:connectivity} and \ref{thm:lifetime}, as detailed below.

We take $\C$ as the circuit in Thm.\ref{thm:connectivity} connecting $\Pi$ to $\rho_{\bbeta}$. Recall that the construction is such that $\C$ is $2\epsilon_\C$-locally reversible with respect to $\Pi$ when the approximation error satisfies $\opnorm{\C[\Pi]-\rho_{\bbeta}}\leq\epsilon_\C$.

Since $\rng\C=\polylog(\poly L/\epsilon_\C)$, we can always find positive constants $c_1, c_2$ such that $\rng\C \leq \log^{c_2} (L^{c_1}/ \epsilon_\C)$ holds for sufficiently large $L$. Letting $\rng\C= L/8$ leads to an error bound
\begin{equation}
    \epsilon_\C \leq \exp(-(L/8)^{c_2^{-1}})\cdot L^{c_1}\leq \exp(-L^{c_2^{-1}+\epsilon})
\end{equation}
where, for any positive $\epsilon$, the second inequality holds when $L$ is large enough. Now Theorem 3 straightforwardly follows from Theorem 2. We remark that in our construction $c_2$ is strictly larger than $1$; therefore, the error term in Thm.~3 follows a stretched-exponential decay in $L$.

\section{Continuous-time construction for commuting Hamiltonians}
In this section, we show that for the special case of commuting Hamiltonians (\ie terms in $\bh$ commute with each other), we can relax the stable clustering condition to the clustering condition only (\ie Def.\ref{def:stable_clustering} with $\delta=0$) and obtain a continuous-time Lindbladian evolution connecting two Gibbs states within the same phase. 

\begin{theorem}\label{thm:cont_connectivity}
Suppose terms in $\bh=\{h_Z\}_Z$ are commuting.  Let $\{\bbeta(s),s\in[0, 1]\}$ be a single-parameter family of couplings satisfying $|\partial_s\beta_{s,Z}|\leq 1$ for all $s$ and $Z$, and $\{\rho_s\propto\exp(-\bbeta_s\cdot\bh)\}_s$ be the corresponding family of Gibbs states. Then, for any $r\geq0$, there exists an $s$-dependent Lindbladian $\mL_s=\sum_Z \mL_{s,Z}$ such that:
\begin{equation}
    \opnorm{\partial_s\rho_s - \mL_s[\rho_s]}_1 \leq {\sum}_Z \cov^{\A}_{\rho_s}(Z, \overline{Z_{+r}})
\end{equation}
where $Z_{+r}$ denotes lattice sites whose distance to $Z$ is no more than $r$. Each $\mL_{s, Z}$ acts on $Z_{+(r+R)}$ and satisfies $\opnorm{\mL_{s, Z}}_{1\to1}\leq 4$. 
\end{theorem}
{According to the theorem, we notice that if $(\xi, f)$-clustering as defined in Def.\ref{def:clustering} is satisfied by each of the $\rho_s$ along the path $s\in[0,1]$, then the Lindbladian evolution approximately transforms $\rho_0$ into $\rho_1$, with an error bounded as:
\begin{equation}
        \epsilon_\mL :=\opnorm{\rho_1 -\mathcal{T}e^{\int \mL ds}[\rho_0]}_1 \leq \poly(L) e^{-r/\xi}
\end{equation}
which can be rewritten as a bound on the range $r$ of the Lindbladian:
\begin{equation}
    r\leq\xi\cdot\log\left(\frac{\poly(L)}{\epsilon_{\mL}}\right).
\end{equation}
}

We now prove Theorem \ref{thm:cont_connectivity}. Expanding the derivative of $\rho_s$, we get:
\begin{equation}
\begin{split}
    \partial_s\rho_s 
    &= {\sum}_{Z} (\partial_s\beta_{s, Z}) \cdot \partial_{\beta_Z}\rho_s \\
    &= {\sum}_{Z} (\partial_s\beta_{s, Z}) \cdot \rho_s\cdot(\braket{h_Z}_{s}-h_Z).
\end{split}
\end{equation}
where $\braket{h_Z}_s =\tr(h_Z\rho_s)$. We assume $\partial_s\beta_{s, Z}\geq 0$ for each $Z$ without loss of generality (otherwise we can redefine $h_Z$ as $-h_Z$).

{We make the following key observation: each term of the derivative—e.g., the one corresponding to $h_Z$—is proportional to the difference between two Gibbs states whose Hamiltonians differ only on $Z$:}
\begin{equation}
\begin{split}
&\rho_s\cdot(\braket{h_Z}_{s}-h_Z) = (\lambda_Z-\braket{h_Z}_s)\cdot\left(\tilde\rho_{s, Z}-\rho_s\right)\\
&{\rm where}\quad\tilde\rho_{s, Z} :=\frac{\lambda_Z-h_Z}{\lambda_Z-\braket{h_Z}_s}\rho_s
\end{split}
\end{equation}
Here $\lambda_Z$ denotes the largest eigenvalue of $h_Z$. It is straightforward to check that $\tilde\rho_{s, Z}$ is a valid density matrix and can be viewed as a Gibbs state corresponding to the Hamiltonian:
\begin{equation}
    \tilde H_{s, Z} = \bbeta_s\cdot\bh-\log\left(\lambda_Z-h_Z + 0_+\right)
\end{equation}
which is also commuting. 

{This observation motivates defining the Lindbladian as:
\begin{equation}
\begin{split}
    \mL_s &= {\sum}_{Z}\mL_{s, Z}\quad{\rm where}\\
    \mL_{s, Z} &:= (\partial_s\beta_{s, Z})\cdot(\lambda_Z-\braket{h_Z}_s) \cdot(\M_{s, Z}-\mathcal{I})
\end{split}
\end{equation}
with $\mathcal{I}$ being the identity channel and $\M_{s, Z}$ a channel to be specified soon.} It is straightforward to check that $\mL_{s, Z}$ is a valid Lindbladian, \ie $e^{t\mL_{s,Z}}$ is a quantum channel for any $t$. Further, since $|\partial_s\beta_{s, Z}|\leq1$ and $|\lambda_Z-\braket{h_Z}_s|\leq 2\opnorm{h_Z}\leq2$, we have:
\begin{equation}
    \opnorm{\mL_{s,Z}}_{1\to 1} 
    \leq 2\opnorm{\M_{s, Z}-\mathcal{I}}_{1\to1} \leq 4
\end{equation}

{The Lindbladian above leads to an approximation error:
\begin{equation}
\begin{split}
&\opnorm{\partial_s\rho_s - \mL_s[\rho_s]}_1\\
    \leq& {\sum}_Z \opnorm{\partial_s\beta_{s, Z}\partial_{\beta_Z}\rho_s -\mL_{s,Z}[\rho_s]}_1\\
    \leq& {\sum}_Z |\lambda_Z-\braket{h_Z}_s|\cdot\opnorm{\tilde\rho_{s, Z}-\M_{s,Z}[\rho_s]}_1 .
\end{split} 
\end{equation}
Thus it remains to choose $\M_{s,Z}$ that approximately transforms $\rho_{s}$ into $\tilde\rho_{s,Z}$. Since the Hamiltonians of $\rho_{s}$ and $\tilde\rho_{s,Z}$ differ only on $Z$, we can use the same strategy as in Sec.\ref{subsec:local_variation}. Letting $B_1=Z_{+r}\setminus Z$, $B_2=Z_{+(r+R)}\setminus Z_{+r}$, and $C=\overline{Z_{+(r+R)}}$, we define:}
\begin{equation}
    \M_{s,Z}= \R^{\tilde\rho_{s,Z}}_{B_2\to ZB}\circ\Tr_{ZB_1}
\end{equation}
where $B=B_1 B_2$ and $\R^{\tilde\rho_{s,Z}}_{B_2\to ZB}$ is the Petz recovery map. Since $\tilde\rho$ is a commuting Hamiltonian’s Gibbs state and dist$(Z, B_2)\geq R$, we know $I_{\tilde\rho}(Z:B_2|B_1)=0$, which leads to $\tilde\rho_{s,Z}=\R^{\tilde\rho_{s,Z}}_{B_2\to ZB}[\tilde\rho^{B_2 C}_{s,Z}]$. Therefore:
\begin{equation}
    \begin{split}
        &\opnorm{\tilde\rho_{s, Z}-\M_{s,Z}[\rho_s]}_1 
        \leq \opnorm{\rho_s^{B_2 C}-\tilde\rho^{B_2 C}_{s,Z}}_1\\
        &=\sup_{O\in \A_{B_2 C},\ \opnorm{O}=1}\left|\Tr(O\rho_s-O\tilde\rho_{s,Z})\right|\\
        &= \frac{1}{|\lambda_Z - \braket{h_Z}_s|}\sup_{O\in \A _{B_2 C}, \opnorm{O}=1} \left|\braket{O h_Z}_s - \braket{O}_s\braket{h_Z}_s\right|\\
        &\leq \frac{1}{|\lambda_Z - \braket{h_Z}_s|} \cov^{\A}_{\rho_s}(Z, \overline{Z_{+r}}),
    \end{split}
\end{equation}
which completes the proof.

\section{Summary and outlook}
To summarize, in this work we provide a circuit-based characterization of Gibbs states within a thermal phase. We formulate a correlation-decay condition dubbed stable clustering (Def.\ref{def:stable_clustering}), which is expected to be satisfied for Gibbs states within a vast class of thermal phases. We prove that there exists a local quantum channel circuit that transforms one Gibbs state into another, provided they are connected by a path that satisfies stable clustering. For the case of commuting local Hamiltonians, we are able to relax stable clustering to a weaker condition (Def.\ref{def:clustering}) and improve the circuit to a continuous-time evolution.  As an application of the construction, we establish that if a system is in the same thermal phase as a zero-temperature topological code, then it can coherently store quantum information for a macroscopically long time while in contact with the thermal bath.

We end this work with some discussions and open questions:
\begin{itemize}[leftmargin=0.5cm]

    \item Our constructions have room for potential improvements. In light of Theorem \ref{thm:cont_connectivity}, it is natural to ask whether one can relax the stability part of the clustering condition for noncommuting Hamiltonians as well, and obtain a continuous-time Lindbladian. In addition, we also ask whether the transforming circuit/Lindbladian can be made exact ($\epsilon_\C=0$) while still preserving locality. The latter question is motivated by the exact quasi-adiabatic continuation for gapped ground states (see \eg \cite{2010hastings}), which exactly transforms one ground state into another while having a transient Hamiltonian with superpolynomially decaying tails.
        
    \item Rigorously establishing the relationship between (stable) clustering and other characterizations of thermal phases, \eg analyticity of the free energy, remains an open problem. Ref.\ \cite{2019complexzero} shows that analyticity of the free energy at high temperature implies decay of correlations for local observables in quantum Gibbs states. Extending this connection to low temperatures and to correlations of nonlocal observables (as in the annulus tripartition) remains open. It is also desirable to obtain (stable) clustering for important models, especially at low temperatures. In App.\ref{app:24TC}, we provide a physical argument suggesting that stable clustering fails for the 2D toric code but holds for the 4D toric code near the zero-temperature limit. Turning this into a rigorous proof is desirable. 

    \item Our current construction of local interaction variation (Eq.\ref{eq:localM}), which is an erase-and-resample procedure, is a choice made for analytical tractability rather than a unique requirement. A physically more natural alternative is a short-time evolution under a detailed-balanced quasi-local Lindbladian of the type constructed in Ref.~\cite{2025CR}, which has the target state $\tilde\rho$ as a steady state. One expects such a dissipative evolution to smoothly deform $\rho$ into $\tilde\rho$ within $AB_1$ over a short time. Obtaining a rigorous bound on the time required for this smooth deformation to reach a given accuracy, however, remains beyond the reach of currently available techniques, and we leave it as an interesting open question.
    
    \item The circuit construction under additional symmetry constraints deserves further study. When the Hamiltonian has a global ($0$-form) symmetry, a natural question is whether the circuit can be made strongly symmetric—our current construction is not, because a partial trace appears in the local interaction variation [Eq.\eqref{eq:localM}]. In App.\ref{app:symmetrize} we provide evidence that, for onsite $\mathbb{Z}_2$ symmetry, local interaction variation can be made strongly symmetric. Making this statement rigorous, and extending it to continuous symmetries, is desirable. Furthermore, when a continuous symmetry is present and spontaneously broken at low temperatures, Goldstone modes produce algebraic spatial correlations, so stable (exponential) clustering fails. Whether two Gibbs states in the ordered regime can still be connected by a circuit—even without requiring strong symmetry—and how its range scales with the system size remain open questions. 

\end{itemize}

\begin{acknowledgements}
We thank Thiago Bergamaschi, Matthew P.A. Fisher, Tarun Grover, Tim Hsieh, Michael Levin, Yaodong Li, David Long, Benedikt Placke, Tibor Rakovszky, Mehdi Soleimanifar, Charles Stahl, Chong Wang and Cenke Xu for helpful discussions. S.S. was supported by the SITP postdoctoral fellowship at Stanford University. V.K. acknowledges support from the Packard Foundation through a Packard Fellowship in Science and Engineering and the Office of Naval Research Young Investigator Program (ONR YIP) under Award Number N00014-24-1-2098.  R.M. was supported in part by the Simons Investigator grant (990660) and by the Simons Collaboration on Ultra-Quantum Matter, which is a grant from the Simons Foundation (651442). R.M. also acknowledges partial support from grant NSF PHY-2309135 and the Simons Investigator program (C. Xu).
\end{acknowledgements}

\bibliography{Ref.bib} 

\appendix
\onecolumngrid

\section{Proof of Lemma 1}\label{app:LPPL}
Recall the variational definition of the trace norm:
\begin{equation}
    \Vert X\Vert_1 = \sup_{\Vert O\Vert\leq 1}\Tr(X O).
\end{equation}
In our case $X = \rho^{C}-\tilde\rho^{C}$. Since both $\rho^{C}$ and $\tilde\rho^{C}$ belong to the algebra $\A_{C}$, it suffices to consider $O$ chosen from $\A_{C}$.

Assuming that $V$ is supported on $A$ and defining $\rho(s)$ as the Gibbs state of $H(s)$, then for any unit-norm operator $O$ supported on $B_2C$:
\begin{equation}
    \Tr[(\rho(0) -\rho(1)) O] =  -\frac{1}{2} \int_0^1 ds\ \cov_{\rho(s)}(\Phi^s(V), O)+\cov_{\rho(s)}(O, \Phi^s(V)),
    \label{eq:LPPL}
\end{equation}
where $\cov_{\rho} (O_1,  O_2):=\Tr(\rho O_1 O_2)-\Tr(\rho O_1)\Tr(\rho O_2)$ denotes the connected correlation in the state $\rho$, while $\Phi^{s}(V)$ is the quantum belief-propagation operator defined as~\cite{hastings2007quantum}:
\begin{equation}
\Phi^{s}(V) := \int_{-\infty}^{\infty} dt \, f(t) \, e^{-itH(s)} \, V \, e^{itH(s)}
\end{equation}
for some exponentially decaying $L_1$ function $f(t)$ (see Eq.(26) in Ref.\cite{2025LPPL} for an explicit expression).

The operator $\Phi_\beta^{\,s}(V)$ has the following properties, crucial for us:
\begin{enumerate}
    \item $\Vert\Phi^s(V)\Vert\leq\Vert V\Vert$
    \item $\Phi^s(V)$ belongs to $\A$, the algebra generated by local interaction terms $\bh$.
    \item  $\Phi^{s}(V)$ can be approximated by another operator $\Phi^{\,s,l}(V)$ supported strictly in the $l$-neighborhood of $A$, denoted by $A_{+l}:=\{x\in\Lambda\mid d(x,A)\le l \}$, up to an error \cite{2023Alhambra,2025LPPL}
\begin{equation}
    \Vert \Phi^s_\beta(V) - \Phi^{s,l}_\beta(V) \Vert \leq \Vert V \Vert\zeta_{\rm QBP}(A,l),
\end{equation}
where $\zeta_{\mathrm{QBP}}(A,l)$ decays exponentially when $H(s)$ is a short-range Hamiltonian,
\begin{equation}
    \zeta_{\rm QBP}(A,l) \leq 6|A|e^{-\frac{1}{1+v_{LR}/\pi}l}.
\end{equation}
Here $v_{LR}$ is the maximum Lieb–Robinson velocity of the Hamiltonians $H(s)_{s\in[0,1]}$. 
\end{enumerate}

Choosing $l<d(A,C)$, we obtain
\begin{equation}
\begin{aligned}
    |\Tr[(\rho(0) -\rho(1)) O] |
    &\leq  \int_0^1 ds\ \cov_{\rho(s)}(\Phi^{s,l}(V), O) +\opnorm{V}\cdot \zeta_{\rm QBP}(A,l)\\
    &\leq \opnorm{V} \left(\sup_{s\in[0,1]}\cov^\A_{\rho(s)}(A_{+l},C) + \zeta_{\rm QBP}(A,l)\right)\\
    &\leq \opnorm{V}\left(\sup_{s \in [0, 1]} \cov^\A_{\rho(s)}(A_{+l}; C) +  6|A| e^{-\frac{1}{1+v/\pi}l}\right)
\end{aligned}
\end{equation}

\section{Proof of Lemma 3}\label{app:recovery}
For any state $\rho$ on $\H_A \otimes \H_B  \otimes \H_C$, there exists a completely positive trace-preserving (CPTP) map $\R_{B\to BC}: \mathcal{B}(\H_B) \mapsto \mathcal{B}(\H_B\otimes \H_C) $, such that \cite{2015Petz,2015JRWW} 
\begin{equation} 
I(A:C|B)_\rho \geq -2\log_2 F(\rho,\R_{B\to BC}(\rho_{AB})), \end{equation} 
where $F(\cdot,\cdot)$ denotes the fidelity between two states, and 
\begin{equation} 
\R_{B\to BC}(\cdot):=\int_{-\infty}^\infty dt \beta_0(t)\R^{t}_{\rho^{BC},\Tr_C}(\cdot) 
\label{eq:recoverc}
\end{equation} 
is the rotated Petz recovery map. Here, $\beta_0(t)=\frac{\pi}{2}(\cosh(\pi t)+1)^{-1}$, and for any reference state $\sigma$ and quantum channel $\E$, we define 
\begin{equation} 
\R^{t}_{\sigma,\E}(\cdot) := \sigma^{\frac{1}{2}-it}\E^\dagger\Bigl[\E(\sigma)^{-\frac{1}{2}+it}(\cdot)\E(\sigma)^{-\frac{1}{2}-it}\Bigr]\sigma^{\frac{1}{2}+it}. \label{eq:rotatedPetz} 
\end{equation}
Furthermore, using the Fuchs–van de Graaf inequality,
\begin{equation}
    1-F(\rho,\sigma) \geq \frac{1}{4} \Vert \rho-\sigma\Vert_1^2,
\end{equation}
we deduce that
\begin{equation}
    I(A:C|B)_\rho \geq \frac{1}{4 \ln2} \Vert \rho - \R_{B\to BC}(\rho^{AB}) \Vert_1^2.
\end{equation}

\section{Symmetric local interaction variation: example of Ising symmetry}
\label{app:symmetrize}
In this appendix we discuss conditions under which local interaction variation can be made to respect global symmetries present in local interaction terms $\bh$. We focus on the special case where $G=\{\identity, g=\prod_i X_i\}$ is the global Ising symmetry. We assume each $h_X$ commutes with $g$, \ie $[h_X, g]=0$ for all $X$. This implies any Gibbs state $\rho_\bbeta$ is weakly symmetric under the Ising symmetry, \ie $g\rho_\bbeta g^{-1} = \rho_\bbeta$.

As we recall, local Hamiltonian variation Eq.\eqref{eq:localM} is achieved by composing $\Tr_{AB_1}$ and a recovery map. It is mathematically equivalent to replace the partial trace with a complete depolarizing channel, defined as $\D_{AB_1}[\rho] = 2^{-|AB_1|}\Tr_{AB_1}[\rho]\otimes\identity_{AB_1}$. Therefore $\M$ can be rewritten as:
\begin{equation}
    \M[\rho_\bbeta] = \R_{B_2\to AB}\circ\D_{AB_1}[\rho_\bbeta].
\end{equation}

The recovery map $\R_{B_2\to AB}$ is strongly $G$-symmetric, as shown below:
\begin{equation}
    g\rho^X_\bbeta g^{-1} = g_{X}\rho^X_\bbeta g_X^{-1}=g_{X}\Tr_{\bar X}[g_{\bar X}\rho_{\bbeta}g^{-1}_{\bar X}]g_X^{-1}=\Tr_{\bar X}[g\rho_{\bbeta}g^{-1}]=\rho^X_\bbeta.
\end{equation}
In the following we discuss when the complete depolarization channel $\D_{AB_1}$'s action on $\rho_\bbeta$ can be simulated by a strongly symmetric channel.

We consider performing complete depolarization in different symmetry sectors separately, as detailed below. For any region $X$, the state $\rho_{\bbeta}$ can be decomposed according to the eigenvalue of $g_X := \prod_{i\in X} X_i$ upon both left and right actions:
\begin{equation}
    \rho_\bbeta = \sum_{s_1, s_2\in \{1, -1\}} \rho_{s_1 s_2} 
\end{equation}
where each $\rho_{s_1 s_2}=P_{s_1}\rho P_{s_2}$ with $P_s$ being the projector to the subspace of $g_X$ with eigenvalue $s$. This can equivalently be understood as the block form of $\rho_\bbeta$ in terms of the $\pm 1$ eigenspaces of $g_X$. The symmetrized depolarization channel is defined as:
\begin{equation}
    \D_X^{\rm sym}[\rho] = \mathbb E_{U_+, U_-} \left[ (U_+\oplus U_-)\rho (U_+\oplus U_-)^\dagger\right] = 
    2^{-|X|+1}(\Tr_{X}(\rho_{++})\otimes P_+ +\Tr_X(\rho_{--})\otimes P_-),
\end{equation}
where $U_+$ and $U_-$ are Haar random unitaries supported on the $\pm 1$ eigenspaces of $g_X$, respectively. In comparison, the complete depolarization channel's action on $\rho_\bbeta$ is given by:
\begin{equation}
    \D_{X}[\rho_\bbeta] = 2^{-|X|}\Tr_{X}(\rho_{++}+\rho_{--})\otimes (P_{+}+P_-).
\end{equation}

The trace distance between the two states is given by (noting $g_X=P_+-P_-$):
\begin{equation}
\begin{split}
    \opnorm{\D^{\rm sym}_X[\rho_\bbeta] - \D_X[\rho_\bbeta]}_1
    =& 2^{-|X|}\opnorm{\Tr_X(\rho_{++})\otimes g_X - \Tr_X(\rho_{--})\otimes g_X}_1\\
    =& \opnorm{\Tr_X(\rho_{++})- \Tr_X(\rho_{--})}_1\\
    =& \opnorm{\Tr_X(\rho_\bbeta P_+)- \Tr_X(\rho_\bbeta P_-)}_1\\
    =& \opnorm{\Tr_X(g_X\rho_{\bbeta})}_1 
\end{split}
\end{equation}
Physically, the expectation value of the symmetry generator restricted to $X$ can be interpreted as a disorder parameter for the \emph{strong} Ising symmetry. Because the Gibbs state $\rho_\bbeta$ lacks a strong Ising symmetry from the outset, this disorder parameter is expected to decay exponentially with the volume $|X|$—in fact, for classical Gibbs states it is exactly zero. Consequently, we expect our local interaction-variation channel can be arranged to respect an on-site Ising symmetry, up to errors that are exponentially small in the volume of $AB_1$ in Fig.~\ref{fig:temperaturevariation}.

\section{Derivation of Eq.\eqref{eq:totalerror}}
\label{app:totalerror}
We introduce the shorthand notation $\G_{\geq t}=\M^{(N)}\circ...\circ\M^{(t)}$. We then have:
\begin{equation}
    \begin{split}
        &\opnorm{\M^{(N)}\circ...\circ \M^{(1)}(\rho_{\bbeta}) - \rho_{\bbeta+\Delta\bbeta}}_1 \\
        \leq
        &\opnorm{\sum_{i=0}^{N-1} \left(\G_{\geq (i+1)}[\rho^{(i+1)}] - \G_{\geq i}[\rho^{(i)}]\right)}_1 \\
        \leq 
        & \sum_{i=0}^{N-1} \opnorm{\G_{\geq (i+1)}[\rho^{(i+1)}] - \G_{\geq i}[\rho^{(i)}]}_1 \\
        \leq 
        & \sum_{i=0}^{N-1} \opnorm{\rho^{(i+1)} - \M^{(i)}[\rho^{(i)}]}_1\\
    \end{split}
\end{equation}
where the last inequality uses the monotonicity of the trace norm under the action of a quantum channel.

\section{Proof of local reversibility}
\label{app:Clocalrev}
Let $\M^{(m)}$ be a gate in the $t$-th layer of the constructed circuit $\C$. One can find an index $n\leq m$, such that $\M^{(1)},...,\M^{(n)}$ are gates in the first $(t-1)$ layers, and gates $\M^{(n+1)},...,\M^{(m-1)}$ are in the same layer as $\M^{(m)}$ and are in front of $\M^{(m)}$ in the prescribed ordering. 

We use $\C' := \M^{(n)}\circ\cdots\circ\M^{(1)}$ to denote the first $(t-1)$ layers of the circuit. Similarly to Eq.~\eqref{eq:intermediatestates}, we denote $\rho^{(n)} := \rho_{\bbeta + \Delta\bbeta_{n}}$. Then we have:
\begin{equation}
\begin{split} &\opnorm{\tilde\M^{(m)}\circ\M^{(m)}\circ\C'[\rho_\bbeta]-\C'[\rho_\bbeta]}_1\\
\leq
& 2 \Vert \C'[\rho_\bbeta] - \rho^{(n)}\Vert_1 + \Vert \tilde\M^{(m)}\circ\M^{(m)}[\rho^{(n)}]-\rho^{(n)}\Vert_1 \\
\leq & 2\epsilon_\C,
\end{split}
\end{equation}
which establishes the local reversibility of $\mathcal{C}$ with respect to $\rho_{\bbeta}$.

\section{Gapped Ground States}
\label{app:groundstates}

Ref.~\cite{2007Has} showed that for a gapped Hamiltonian whose excitations are point-like quasiparticles, $\rho_{s\bbeta}$ approximates the ground-state density matrix $\rho_{\mathrm{g.s.}}$ within an error $\epsilon$ when $s \ge \log(L/\epsilon)/{\Delta}.$
Here we demonstrate that the same result holds when the excitations reside on higher-dimensional closed manifolds (\eg loops or membranes) with an energy cost proportional to their volume. We expect this assumption on the excitation spectrum to cover a broad class of physically relevant systems.

Consider a Hamiltonian on a $D$-dimensional regular lattice whose excitations live on closed $p$-dimensional surfaces with $p<D$. In the excitation basis, the Gibbs state reads
\begin{equation}
    \rho_{s\bbeta} = \frac{1}{Z} \sum_{\{ \gamma \}}e^{-s E_\gamma} |\gamma \rangle \langle \gamma |,
\end{equation}
where $\gamma$ labels an excitation configuration. As shown in Ref.~\cite{20253DfTC}, the probability that $\rho_{s\bbeta}$ is supported on configurations containing an excitation of volume $V$ is bounded by
\begin{equation}
    \mathrm{P}(V) \leq \Omega_V e^{-s \Delta V},
\end{equation}
where $\Omega_V$ is the number of closed $p$-dimensional surfaces of volume $V$ and $\Delta$ is the energy cost per unit volume. On a regular lattice one has $\Omega_V \leq L^D (c_p)^V$, with $c_p$ the coordination constant (\eg $D-p$ for a hypercubic lattice). Consequently, the total probability of non–ground–state configurations satisfies
\begin{equation}
    \sum_{V>0} \mathrm{P}(V) \leq \frac{c_p L^D e^{-s \Delta}}{1-c_p e^{-s \Delta}}.
\end{equation}
Since the Gibbs state is diagonal in the energy basis, we have
\begin{equation}
    \Vert \rho_{\mathrm{g.s.}} - \rho_{s\bbeta} \Vert_1 = 2 \sum_{V>0} \mathrm{P}(V).
\end{equation}
It follows that choosing $s = \mathcal{O}(\ln(L/\epsilon))$ ensures this trace distance is at most $\epsilon$.

\section{Comparison of 2D and 4D Toric Codes}

\label{app:24TC}
As a simple example, we illustrate in this section that the 2D toric code does not satisfy stable clustering as we approach \(T=0\). To do so, it is sufficient to find a correlation function in the annular-shaped partition that does not decay as in Eq.~\eqref{eq:decayofc}.

\begin{figure}[htb]
  \centering
 
  \resizebox{0.24\columnwidth}{!}{%
    \begin{tikzpicture}[line cap=round]
      \foreach \x in {-1,...,7} {
        \draw[black, thin] (\x,-1) -- (\x,7);
      }
      \foreach \y in {-1,...,7} {
        \draw[black, thin] (-1,\y) -- (7,\y);
      }
      \draw[red, thick] (2,2) rectangle (4,4);
      \draw[blue, thick] (0,0) rectangle (6,6);
    \end{tikzpicture}%
  }
  \caption{}
  \label{fig:squarelattice}
\end{figure}

Let us consider the square lattice in Fig.~\ref{fig:squarelattice}, where qubits reside on the links. The 2D toric code Hamiltonian is
\begin{equation}
    H = -\sum_\Box A_\Box  - \sum_+ B_+,
\end{equation}
where \(A_\Box = \prod_{i\in\Box} Z_i\) and \(B_+ = \prod_{i\in +} X_i\) are the usual plaquette and vertex terms, respectively. Define two loop operators supported on the loops \(\gamma_1\) and \(\gamma_2\) (the red and blue loops in Fig.~\ref{fig:squarelattice}, respectively):
\begin{equation}
    O_1  = \prod_{i\in\gamma_1} Z_i,\quad O_2  = \prod_{i\in\gamma_2} Z_i.
    \label{eq:correlationloops}
\end{equation}
Let us denote the region enclosed by \(\gamma_1\) as \(A\), and that by \(\gamma_2\) as \(B\). We begin with a Gibbs state $\rho_\bbeta$ at uniform inverse temperature $\beta_0$, such that $\rho_\bbeta$ approximates the ground-space state $\rho_{\mathrm{g.s.}}$ within error $\epsilon$. As discussed in Sec.~\ref{app:groundstates}, it suffices to choose $\beta_0 \ge \log\bigl(\mathrm{poly}\,L/\epsilon\bigr)$. We then choose $\Delta \bbeta$ so that in the state $\rho_{\bbeta + \Delta \bbeta}$, all Hamiltonian terms fully supported within region $A$ are at a finite inverse temperature $\beta$, while those supported nontrivially outside \(\gamma_1\) remain unchanged.

We now compute the connected correlation function of the two closed-loop operators $O_1$ and $O_2$ in the state $\rho_{\bbeta+\Delta\bbeta}$. The operators \(O_1\) and \(O_2\) simply measure the parity of $m$ anyons (the plaquette excitations) within regions \(A\) and \(B\), respectively:
\begin{equation}
\begin{split}
    &\Tr(\rho_{\bbeta+\Delta\bbeta} O_1) = \prod_{i\in A} \Tr(\rho_{\bbeta+\Delta\bbeta}\mathrm{P}_i) = (\frac{1-e^{-2\beta}}{1+e^{-2\beta}})^{N_A} = (\tanh\beta)^{N_A},  \\
     &\Tr(\rho_{\bbeta+\Delta\bbeta} O_2) = \prod_{i\in B} \Tr(\rho_{\bbeta+\Delta\bbeta}\mathrm{P}_i) = (\tanh\beta)^{N_A} (\tanh \beta_0)^{N_B-N_A},
\end{split}  
\end{equation}
where $\mathrm{P}_i$ measures the parity of the $i$-th plaquette, and $N_A$ is the total number of plaquettes in $A$, and similarly for $N_B$. On the other hand, \(O_1 O_2\) measures the parity of the \(m\) anyons within \(B\setminus A\):
\begin{equation}
    \Tr(\rho_{\bbeta+\Delta\bbeta} O_1 O_2) = (\tanh\beta_0)^{N_B-N_A},
\end{equation}
For the state $\rho_\bbeta$ to approximate the ground space within $\epsilon$, we must have 
\begin{equation}
    (\tanh\beta_0)^{N_B-N_A}=\Tr(\rho_{\bbeta+\Delta\bbeta} O_1 O_2) = \Tr(\rho_{\bbeta} O_1 O_2)>1-\epsilon.
\end{equation}
Therefore, 
\begin{equation}
    \Tr(\rho_{\bbeta+\Delta\bbeta} O_1 O_2) -\Tr(\rho_{\bbeta+\Delta\bbeta} O_1)\Tr(\rho_{\bbeta+\Delta\bbeta} O_2)  > (1-\epsilon)[ 1-(\tanh\beta)^{2N_A}] >0,
\end{equation}
This nonzero connected correlation demonstrates that $(\delta,\xi)$-stable clustering fails in the 2D toric code when the $\delta$-ball in parameter space attempts to approach the zero-temperature limit from a finite temperature.

In contrast, in the 4D toric code the analogous connected correlation vanishes. On a 4D hypercubic lattice, the Hamiltonian is
\begin{equation}
    H_{4D} = -\sum_s A_s - \sum_c B_c,
\end{equation}
where qubits reside on two-dimensional plaquettes, and
\begin{equation}
    A_s = \prod_{i\in\partial s} Z_i,\quad B_c = \prod_{\partial i \ni c} X_i,
\end{equation}
are the stabilizer terms associated with three-dimensional cubes (denoted by $\{s\}$) and one-dimensional edges (denoted by $\{c\}$), respectively. In the annular partition, the region $A$ is the four-dimensional solid ball $B^4$. We then consider the state $\rho_{\bbeta + \Delta \bbeta}$ in which all stabilizers acting nontrivially outside $A$ are in the zero-temperature limit, while those supported entirely within $A$ remain at finite inverse temperature $\beta$.

Let $\bar{A} = \Lambda\setminus A\simeq S^3$ denote the complement of region $A$ on the lattice. As in the main text, let $\mathcal{A}$ be the algebra generated by all Hamiltonian terms, and define $\mathcal{A}_{\bar A} = \Tr_A \mathcal{A}$. Moreover, let $\mathcal{B}_{\bar A}$ be the subalgebra generated by Hamiltonian terms supported entirely in $\bar{A}$. Evidently, $\mathcal{B}_{\bar A} \subset \mathcal{A}_{\bar A}$. Compared with the 2D case, the 4D case has a crucial distinction: In the case where $\bar A$ is isomorphic to $S^3$, one can verify that $\mathcal B_{\bar A} = \mathcal A_{\bar A}$, i.e., any product of stabilizers acting nontrivially in $A$ cannot yield an operator whose support lies entirely in $\bar A$ that is \emph{not} itself a product of stabilizers contained in $\bar A$.\footnote{In 2D, the operator $O_2$ in Eq.~\eqref{eq:correlationloops} lies in $\A_{\bar A}$ but not in $B_{\bar A}$.
} Consequently, to examine the covariance between $A$ and $\bar A$ as defined in Eq.\eqref{eq:covariance}, it suffices to consider the correlation
\begin{equation}
  \Tr\bigl(\rho_{\bbeta + \Delta\bbeta}\,O_1 O_2\bigr)
  - \Tr\bigl(\rho_{\bbeta + \Delta\bbeta}\,O_1\bigr)\,
    \Tr\bigl(\rho_{\bbeta + \Delta\bbeta}\,O_2\bigr),
\end{equation}
where $O_1\in \mathcal A_{A}$ and $O_2\in \mathcal A_{\bar A} = \mathcal B_{\bar A}$.

Now, by taking the temperature of all terms acting nontrivially outside $A$ to zero, the state $\rho_{\bbeta+\Delta\bbeta}$ expels all excitations from $\bar A$. More precisely, if $\beta_0$ is chosen such that $\|\rho_\bbeta - \rho_{\mathrm{g.s.}}\|_1 \le \epsilon$,
then by the same argument as in Sec.~\ref{app:groundstates}, all stabilizers supported in $\bar A$ have eigenvalue $1$ up to an error $\epsilon$, so
\begin{equation}
  \Tr\bigl(\rho_{\bbeta+\Delta \bbeta} O_1 O_2\bigr)
  \eq{\epsilon}
  \Tr\bigl(\rho_{\bbeta+\Delta \bbeta} O_1\bigr)
  \eq{\epsilon}
  \Tr\bigl(\rho_{\bbeta+\Delta \bbeta} O_1\bigr)\,
  \Tr\bigl(\rho_{\bbeta+\Delta \bbeta} O_2\bigr).
\end{equation}
where $O_1\in \A_A$ and $O_2\in \mathcal B_{\bar A}$, both of operator norm $1$, indicating that the annular-shaped covariance vanishes in the state $\rho_{\bbeta+\Delta \bbeta}$.

We emphasize that the latter part does not provide a direct proof that stable clustering holds for the 4D toric code in the low-temperature phase; rather, it offers physical understanding into the differences between the 2D and 4D cases.

\section{Local indistinguishability of quantum codes}\label{app:clear_lemma}
Suppose $\Pi$ is a code space projector of a diameter-$\ell$ quantum code, with codespace dimension $2^k$. We introduce a reference system $R$ and have it maximally entangled with the code space, such that the total state is
\begin{equation}
    \ket{\Psi_{RQ}} =2^{-k/2}\sum_{a=1}^{2^k} \ket{a}_R\otimes \ket{\tilde a}_Q,
\end{equation}
where $\{\ket{a}\}$ is a basis of $R$, and $\{\ket{\tilde a}\}$ is a basis of the code space $\Pi$. We have used $Q$ to denote the physical qubits supporting the code space $\Pi$. Let $A\subset Q$ be a region with diameter less than $\ell$. Then, by the definition of topological codes, the code states $\{\ket{\tilde a}_Q\}$ are indistinguishable on $A$, leading to:
\begin{equation}
    I_{\ket\Psi}(R:A) = 0 
\end{equation}
Following the theorem in Ref.~\cite{hayden2004structure}, the Hilbert space $\H_{\overline A}$ admits a decomposition $\H_{\overline A}=(\H_X\otimes\H_Y)\otimes \H'$ under which the state $\ket{\Psi_{RQ}}$ takes the form
\begin{equation}
    \ket{\Psi_{RQ}} = 2^{-k/2}\sum_{a=1}^{2^k} \ket{a}_R\otimes \ket{\tilde{a}'}_{R X}\otimes\ket{\omega_0}_{Y A}
\end{equation}
such that $\ket{\omega_0}_{Y A}$ is a fixed state on $Y\cup A$ independent of $a$. Now, for any two code states $\rho_Q$ and $\sigma_Q$ within the codespace $\Pi$, under the decomposition above, they must be of the form
\begin{equation}
    \rho_{Q} = \rho'_{X}\otimes \ket{\omega_0}\bra{\omega_0}_{Y A}\quad\quad  \sigma_{Q} = \sigma'_{X}\otimes \ket{\omega_0}\bra{\omega_0}_{Y A}.
\end{equation}
Letting $\D[\cdot] = \Tr_{X}(\cdot)\rho'_{X}$ and $\D'[\cdot] = \Tr_{X}(\cdot)\sigma'_{X}$, we have $\D[\sigma]=\rho$ and $\D'[\rho]=\sigma$. Since $\D$ and $\D'$ are local operations supported on $\overline A$, $\rho$ and $\sigma$ are $(\ell, 0)$-LI.

\section{Local reversibility and local indistinguishability}\label{app:LR_LI}
A few notations for this appendix: $\rho \eq{\delta}\sigma$ means $\opnorm{\rho-\sigma}_1\leq \delta$. $\G=\prod_X\E_X$ is used to denote one layer of a channel circuit. The gate $\E_X$ acts within a simply connected region $X$. Different $X$s do not overlap. $\rng(\G) := \sup_{X}\diam(X)$. $|\G|$ means the number of gates in $\G$.

\noindent{\bf Def:} A layer $\G=\prod_X\E_X$ is locally reversible (L. R.) with respect to $\rho$ if, for each $X$, there exists a channel gate $\tilde\E_X$ supported within $X$ such that $\tilde\E_X\circ\E_X[\rho]\eq{\delta}\rho$. 

\noindent{\bf Claim:} If $\G$ is $\delta$-L.R. with respect to $\rho$, then $\tilde\G\circ\G[\rho]\eq{|\G|\cdot\delta}\rho$.
The proof is straightforward.

\noindent{\bf Def:} $\rho$ and $\sigma$ are $(r, \delta)$ locally indistinguishable (L.I.) if, for any simply connected $A$ with $\diam(A)<r$, there exist channels $\D, \D'$ supported within $\overline A$ such that $\D[\rho]\eq{\delta}\sigma,\ \D'[\sigma]\eq{\delta}\rho$.

For instance, any two codeword states $\sigma,\rho$ of the toric code on an $L\times L$ torus are $(L/2, 0)$-L.I.

\noindent{\bf Claim:} If $\rho$ and $\sigma$ are $(r, \delta_{LI})$-L.I., and $\G$ is $\delta_{LR}$-L.R. with respect to $\sigma$, then: 
\begin{enumerate}
    \item $\G$ is $(2\delta_{LI}+\delta_{LR})$-L.R. with respect to $\rho$.
    \item {$\G[\rho]$ and $\G[\sigma]$ are $(r-2\rng\G,(2|\G|+1)\delta_{LI} +|\G|\delta_{LR})$-L.I.}
\end{enumerate}
Proof of the first claim is straightforward. For each $X$,
\begin{equation}
    \tilde\E_X\circ\E_X[\rho]
    \eq{\delta_{LI}}
    \tilde\E_X\circ\E_X\circ\D[\sigma]
    =\D\circ\tilde\E_X\circ\E_X[\sigma]
    \eq{\delta_{LR}}
    \D[\sigma]
    \eq{\delta_{LI}}
    \rho
\end{equation}
where $\D$ is chosen to be within $\overline X$.

To prove the second claim, let $A$ be any region with $\diam(A)\leq r-2\rng(\G)$. Let $B$ be a width-$\rng \G$ buffer around $A$, and let $C=\overline{AB}$. Since $\diam(AB)\leq r$, there exists $\D$ in $C$ such that $\D[\rho]\eq{\delta_{LI}}\sigma$. 

Let $\G=\G_C\G_{\bar C}$ where $\G_C$ contains all gates having overlap with $C$, and $\G_{\bar C}$ contains the remaining gates. Then $\G_{\bar C}$ is supported within $AB$. {Let $\D^\G := \G_C \D\tilde\G_C$ and $\D'^\G := \G_C \D' \tilde\G_C$, both acting on $BC$. Then
\begin{equation}
\begin{split}
&\D'^\G\circ \G[\sigma] = \G_C \D' \tilde\G_C \G_C \G_{\bar C}[\sigma]
    \eq{|\G_C|\cdot\delta_{LR}}
    \G_C \D' \G_{\bar C}[\sigma]\eq{\delta_{LI}}\G[\rho] \\
&\D^\G\circ \G[\rho] = \G_C \D \tilde\G_C \G_C \G_{\bar C}[\rho]
    \eq{|\G_C|\cdot(2\delta_{LI}+\delta_{LR})}
    \G_C \D \G_{\bar C}[\rho]\eq{\delta_{LI}}\G[\sigma], 
\end{split}
\label{eq:onelayerLI}
\end{equation}
which completes the proof.

We now generalize the result to circuits with multiple layers. Consider a circuit $\C = \G_T\G_{T-1}...\G_1$ such that each layer $\G_t$ is $\delta$-LR with respect to $\C_{t-1}[\sigma]$, where $\C_t := \G_t \G_{t-1}\cdots \G_1$ (i.e., the first $t$ layers of $\C$). Further, let $\rho$ and $\sigma$ be $(r, 0)$-LI. Consider any region $A$ with $\diam(A)\le r-2\rng{\C}$, and let $B$ be a width-$\rng{\C}$ buffer region that shields $A$ from $C=\overline{AB}$. Analogously, define channels $\D_t := \C_{t,C} \D \tilde\C_{t,C}$ and $\D_t' := \C_{t,C} \D' \tilde\C_{t,C}$ ($t=0,...,T$), where $\C_{t,C}$ contains all gates in $\C_t$ that overlap with the \emph{light cone} of region $C$, with $\tilde\C_{t,C}$ defined analogously so that it consists of the local inverses of the gates in $\C_{t,C}$. Since $\D$ and $\D'$ are supported in region $C$, for $t\le T$, $\D_t$ and $\D_t'$ have support outside region $A$. Since $\C$ is $\delta$-LR with respect to $\sigma$, it follows that for $t\leq T$,
\begin{equation}
   \D_{t}' \circ \C_{t}[\sigma] \eq{|\C_{t,C}|\delta} \C_t\circ \D'[\sigma] = \C_t[\rho]. 
\end{equation}
As a result, a gate $\E_{t,X}$ in $\G_t$ that is supported on $A$ is $(2|\C_{t-1,C}|+1)\delta$-L.R. with respect to $\C_{t-1}[\rho]$:
\begin{equation}
    \tilde \E_{t,X} \circ \E_{t,X}\circ \C_{t-1}[\rho] \eq{|\C_{t-1,C}|\delta} \tilde \E_{t,X} \circ \E_{t,X}\circ \D_{t-1}' \circ \C_{t-1}[\sigma] \eq{\delta} \D_{t-1}' \circ \C_{t-1}[\sigma] \eq{|\C_{t-1,C}|\delta}  \C_{t-1}[\rho].
\end{equation}
Consequently,
\begin{equation}
    \D_T\circ C_T[\rho] \eq{\epsilon} C_T[\sigma]\quad\quad\epsilon=|\C_{T,C}|\cdot(2|\C_{T,C}|+1)
\end{equation}
which implies that $\C[\rho]$ and $\C[\sigma]$ are $(r-2\rng \C, 2|\C|^2+|\C|)$-L.I.. This completes the proof of Lemma \ref{lemma:LI_LR}.
}


\section{Derivation of Eq.\eqref{eq:mem_bound}}\label{app:mem_bound}
Here we prove the following claim: if a Lindbladian $\mL$ and a state $\rho$ satisfy $\opnorm{\mathcal{L}[\rho]}_1\leq\epsilon$, then $\opnorm{e^{t\mL}[\rho]-\rho}_1\leq t\epsilon$. For any $\tau$ we have:
\begin{equation}
\begin{aligned}
\opnorm{e^{\tau\mathcal{L}(\rho)} - \rho}_1 &= \opnorm{\sum_{n=1}^{\infty} \frac{\tau^n}{n!} \mathcal{L}^n[\mathcal{L}(\rho)]}_1 \\
&\leq \sum_{n=1}^{\infty} \frac{\tau^n}{n!} \opnorm{\mathcal{L}^n[\mathcal{L}(\rho)]}_1 \\
&\leq \sum_{n=1}^{\infty} \frac{\tau^n}{n!} \opnorm{\mathcal{L}}_{1}^{n-1} \cdot \epsilon \\
&\leq \sum_{n=1}^{\infty} \frac{\tau^n}{n!} \opnorm{\mathcal{L}}_{1}^n \cdot \frac{\epsilon}{\opnorm{\mathcal{L}}_{1}} \\
&= \left(e^{\tau\opnorm{\mathcal{L}}_1} - 1\right) \frac{\epsilon}{\opnorm{\mathcal{L}}_1}
\end{aligned}
\end{equation}

Suppose $t=m\cdot \tau$ with $m$ being an integer. Then we have:
\begin{equation}
\begin{aligned}
    \opnorm{e^{\tau\mL}(\rho) - \rho}_1
    \leq
    &\sum_{k=0}^{k=m-1}\opnorm{e^{(k+1)\tau\mL}[\rho]-e^{k\tau\mL}[\rho]}_1\\
    \leq
    &\sum_{k=0}^{k=m-1}\opnorm{e^{\tau\mL}[\rho]-\rho}_1\\
    \leq
   &\left(e^{t\opnorm{\mathcal{L}}_1/m} - 1\right) \frac{m\cdot\epsilon}{\opnorm{\mathcal{L}}_1}\\
   \lim_{m\to\infty}=
   &\epsilon\cdot t
\end{aligned}
\end{equation}

\end{document}